%% file: 0main.tex
\documentclass[15pt,a4paper]{article}
\usepackage[ ]{natbib} 
\usepackage[english]{babel}
\usepackage[utf8]{inputenc}
\usepackage[T1]{fontenc}
\usepackage{amsmath,amssymb,amsfonts,amsthm}
\usepackage[fleqn,tbtags]{mathtools}
\usepackage{dsfont} 
\usepackage{authblk}
\usepackage{graphicx}
\usepackage{xcolor}
\usepackage{float}
\usepackage{geometry}
\usepackage{booktabs}
\usepackage{multirow}
\usepackage{xr-hyper} 
\usepackage{hyperref}
\usepackage{array}
\usepackage[labelfont=bf]{caption}
\usepackage{floatpag} 
\floatpagestyle{empty} 
\usepackage{setspace} 
\usepackage{mathpazo} 

\usepackage[normalem]{ulem} 

\usepackage[ruled,vlined]{algorithm2e}

\DeclareMathOperator{\Var}{\mathbb{V}\text{ar}}

\DeclareMathOperator{\Tra}{Trace}

\DeclareMathOperator{\E}{\mathbb{E}}

\DeclareMathOperator{\diag}{diag}

\newcommand{\norm}[1]{\left\lVert#1\right\rVert}

\newcommand{\transpose}{{}^{\text{\sffamily T}}}

\DeclareMathOperator{\AR}{AR(1)} 
\DeclareMathOperator{\Exc}{Exc} 
\DeclareMathOperator{\Ind}{Ind}
\DeclareMathOperator{\Reff}{Reff}
\DeclareMathOperator{\QIC}{QIC}
\DeclareMathOperator{\asymQIC}{asymQIC}
\DeclareMathOperator{\CIC}{CIC}
\DeclareMathOperator{\asymCIC}{asymCIC}
\DeclareMathOperator{\SE}{SE}
\DeclareMathOperator{\SD}{SD}
\DeclareMathOperator{\GEEE}{GEEE}
\DeclareMathOperator{\GEE}{GEE}
\DeclareMathOperator{\ER}{ER}
\DeclareMathOperator{\QR}{QR}

\newtheorem{theorem}{Theorem}

\makeatletter
\newcommand*{\addFileDependency}[1]{
  \typeout{(#1)}
  \@addtofilelist{#1}
  \IfFileExists{#1}{}{\typeout{No file #1.}}
}
\makeatother

\title{A new GEE method to account for heteroscedasticity, using asymmetric least-square regressions} 

\author[1,2]{Amadou Barry\footnote{Corresponding author: amadou.barry@mcgill.ca.} }
\author[3]{ Karim Oualkacha}
\author[3]{Arthur Charpentier}

\affil[1]{Departments of Epidemiology, Biostatistics and Occupational Health, McGill University, Montréal, Québec, Canada}
\affil[2]{Lady Davis Institute, Jewish General Hospital, Montréal, Québec, Canada}
\affil[3]{Department of Mathematics and Statistics, Université du Québec à Montréal, Montréal, Québec, Canada}
\date{\today}

\begin{document}
\maketitle

\input{1abstract}
\input{2intro}
\input{3model}

\input{4simulation}
\input{5application}

\input{6conclusion}

\clearpage
\bibliographystyle{apalike}
\bibliography{0main.bib}
\end{document}

%% file: 1abstract.tex
\begin{abstract}
\noindent
Generalized estimating equations $(\GEE)$ are widely used to analyze longitudinal data; however, they are not appropriate for heteroscedastic data, because they only estimate regressor effects on the mean response{\textemdash}and therefore do not account for data heterogeneity. Here, we combine the $\GEE$ with the asymmetric least squares (expectile) regression to derive a new class of estimators, which we call generalized expectile estimating equations $(\GEEE)$. The $\GEEE$ model estimates regressor effects on the expectiles of the response distribution, which provides a detailed view of regressor effects on the entire response distribution. In addition to capturing data heteroscedasticity, the GEEE extends the various working correlation structures to account for within-subject dependence. We derive the asymptotic properties of the $\GEEE$ estimators and propose a robust estimator of its covariance matrix for inference (see our R package, \url{github.com/AmBarry/expectgee}). Our simulations show that the GEEE estimator is non-biased and efficient, and our real data analysis shows it captures heteroscedasticity.
\end{abstract}
{\bf Keywords:} Expectile regression, quantile regression, $\GEE,$ working correlation, cluster data, longitudinal data.


%% file: 2intro.tex
\section{Introduction}\label{Introduction_geee}

Longitudinal data, which is collected at multiple occasions over a period of time, is generally preferred to cross-sectional data. However, statistical analysis of such data is challenging, for two reasons: 1. Within-subject dependence (multiple measurements for the same subject); and 2. Heteroscedasticity. 

Among the available methods for analyzing longitudinal data, generalized estimating equations $(\GEE)$ are undoubtedly one of the most popular~\citep{LiangZeger1986} because they effectively account for within-subject dependence. To develop the $\GEE$ model{\textemdash}which is an extension of the generalized linear model \citep{glm1972} in the longitudinal framework{\textemdash}\citet{LiangZeger1986} relied on the quasi-likelihood concept \citep{Wedderburn1974}, and derived the $\GEE$ estimators as the solution of the estimating equations. As a result, the $\GEE$ model is semi-parametric. They also formally included a working correlation structure in the estimating equations to account for within-subject dependence in the estimation process. 

Among its many attractive qualities, the $\GEE$ model offers a broad class of models for different types of responses and various working correlations to specify the within-subject dependence \citep{HardinJames2003}. Moreover, \citet{LiangZeger1986} have shown that the $\GEE$ estimator is highly efficient even when the true covariance structure is misspecified. Unfortunately, despite all these attractive properties, the $\GEE$ model is inefficient for coping with heteroskedasticity{\textemdash}a common feature of longitudinal data. 

Several past works have tried, unsuccessfully, to improve GEE using quantile regression (QR). \citet{MarinoFarcomeni2015} reviewed all the available studies. They concluded that none of the available research was able to adapt QR to the GEE model, because in each case, it was too challenging to preserve the GEE correlation structure in the QR framework.

More specifically, it is difficult to preserve the random error's covariance structure in the QR framework. For example, \citet{LengZhang2014} showed that when the random noise vector for the repeated measurement from an individual followed an AR(1) structure, the corresponding noise vector for the QR was not AR(1). This is obviously undesirable, because we always want to preserve known aspects of the data in the generalization. 

Consider that we would like to extend the QR covariance to the GEE covariance structure, but it is complex because the former covariance is different from the latter. Indeed, \citet{LengZhang2014} could not preserve the covariance structure because the QR covariance contains a density function, but the GEE covariance does not. This difference makes it hard to extrapolate any assumption from the quantile to the GEE. In addition, it is hard to estimate the QR density function itself, because it is computationally intensive and subject to numerical issues \citep{Chen2004, YinCai2005a, kocherginskyPracticalConfidenceIntervals2005}.

To the best of our knowledge, there is no model that estimates the marginal effect of the regressors on the response distribution, and that naturally generalizes the GEE model's correlation structures. In this paper, we take a new approach to this problem. Instead of attempting to use quantiles, we use expectiles to successfully generalize the working correlation structure of the random error. This lets us model the GEE model in a way that accounts for heteroscedasticity (heterogeneity) in the data.

Since it is well known that quantile models are more robust than expectiles, some may question why we are using expectiles instead of quantiles. Indeed, quantiles are more robust because they generalize the median (whereas expectiles generalize the mean), and are therefore less affected by outliers. However, as mentioned above, all attempts thus far to adapt QR to the GEE model have failed (because they fail to preserve the working correlation structure of the random error). Therefore, we think expectiles are a good alternative.

The word expectile was first coined by \citet{newey1987} in their seminal paper wherein they also introduced the ER. \citet{newey1987} presented a detailed study of this new class of estimator, including its location and scale equivariance, and used the expectile regression (ER) coefficient estimators to test for the presence of heteroscedasticity. \citet{newey1987} also advocated the ER model as an alternative to the QR model introduced earlier by \citet{koenker_regression_1978}; and highlighted some advantages of the $\ER$ over the $\QR.$ In the same period, \citet{Aigner1976} showed that the ER estimators can be derived under a likelihood-based approach using a normal density function with unequal weight on the positive and negative random errors.

Today, there is a growing interest in the $\ER$ topic as demonstrated by the number of publications in the literature. The $\ER$ has been extended to many other classes of models, such as Bayesian \citep{Majumdar2016, Waldmann2016, Xing2017}, nonparametric \citep{Righi2014, YangZou2015}, nonlinear \citep{KimLee2016}, neural network \citep{XuLiuJiang2016, JiangJiang2017, linExpectileNeuralNetworks2020}, support vector machine \citep{Farooq2017}, and splines \citep{SchulzeWaltrup2015}. Moreover, because of its computational advantage, the $\ER$ has now become an important statistical model in high-dimension \citep{zhaoExpectileRegressionAnalyzing2018, zhaoRobustEstimationShrinkage2019, xuElasticnetPenalizedExpectile2020a, panLargeScaleExpectileRegression2020a, panDistributedOptimizationStatistical2020, liaoPenalizedExpectileRegression2019, barryAsymmetricInfluenceMeasure2020}, to name a few. 

In this paper, we consider the ER and propose a Generalized Expectile Estimating Equations $(\GEEE)$ model that retains the attractive properties of the $\GEE$ model, while accounting for heteroskedasticity in the longitudinal data. We derived its asymptotic properties and proposed a heterogeneous consistent and robust estimator of its variance-covariance matrix. We have made a free R package available on GitHub (\url{github.com/AmBarry/expectgee}) to simplify its implementation. Our $\GEEE$ model estimates the regressor effects at the conditional expectile of the response distribution without making any assumption about the random error distribution. In addition, the $\GEEE$ inherits the attractive properties of the classic $\GEE$, in that it provides a highly efficient and consistent estimator even if the true covariance structure is misspecified.

Our main contributions are: 1. Deriving the asymptotic properties of the new GEEE model; 2. Showing that the GEEE model can generalize the various covariance structures available in the GEE framework{\textemdash}in other words, showing that the GEEE model offers a variety of correlation structures to flexibly model the correlation of the within-subject observations; 3. Proposing the variance-covariance matrix for the GEEE estimator; and 4. Generalizing two selection criteria (namely the $\QIC$ and $\CIC$, proposed by \citet{Pan2001} and \citet{cicCriterion2009}, respectively). 

Our model works because in the presence of heteroskedasticity, the GEEE's estimated parameters depend on the expectiles. Thus, by estimating several regression coefficient vectors at different locations of the conditional response distribution, it is possible to study the data's heteroskedasticity in depth. In addition to accounting for the presence of heteroscedasticity in the data, our GEEE model captures regressor effects in the location, scale, and shape of the response distribution. The GEEE model is computationally efficient and easy to implement, and we believe it will be a useful instrument in the toolkit of researchers interested in analyzing longitudinal data.



\begin{center}
\begin{figure}[hbt!]
\includegraphics[width=1\linewidth]{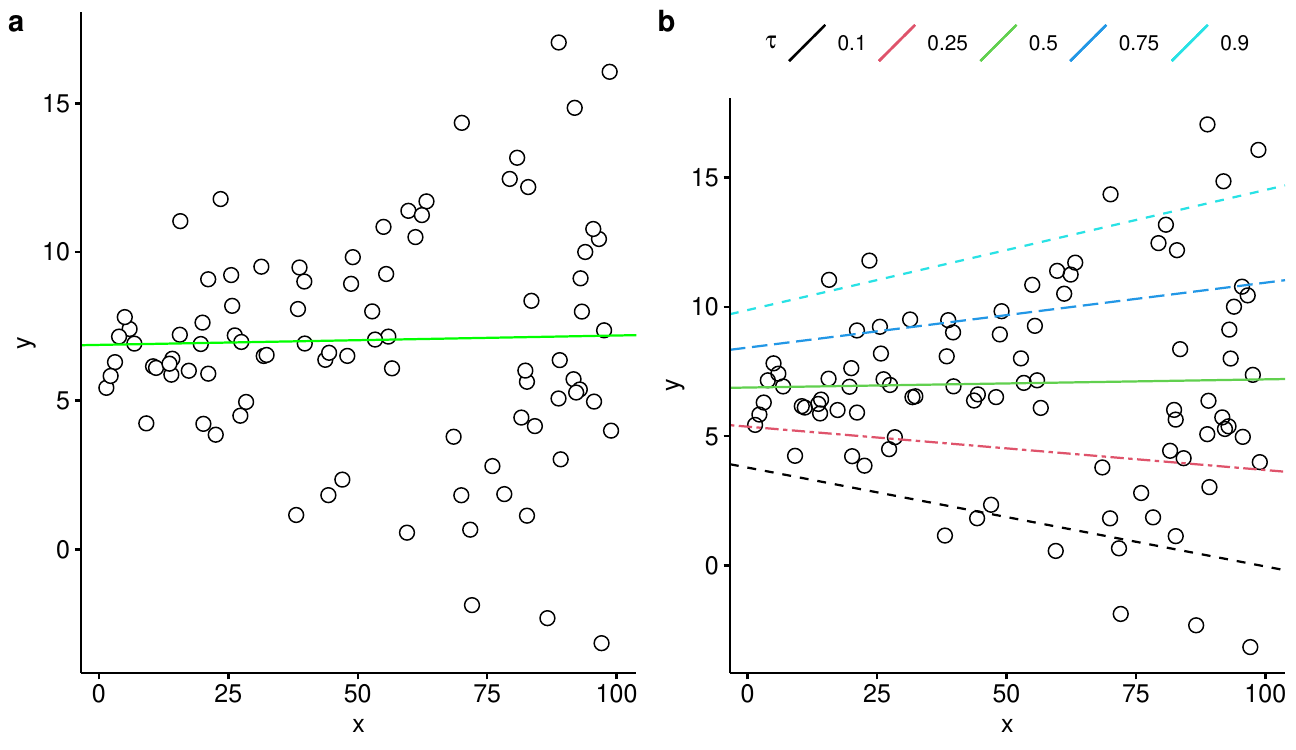}
  \caption{Comparison of the GEE model (left) and GEEE model (right) for a heteroscedastic sample. Figure \textbf{a} shows a GEE mean regression line fitted to the data, and Figure \textbf{b} shows the five GEEE regression lines, $\tau \in (0.1, \ 0.25, \ 0.5, \ 0.75, \ 0.9)$, fitted to the data. The sample $(n=90)$ is generated from a heteroscedastic linear model, $\ y = 6 + 0.025x + \varepsilon,$ where $\varepsilon \sim \mathcal{N}(0, \ \sigma^2)$ and $\sigma=1 + 0.05x.$ }\label{fig:illustration_plot}
\end{figure}
\end{center}

Figure \ref{fig:illustration_plot} highlights the usefulness of our GEEE model. On the left, a classical GEE mean regression is fitted to the data, as indicated by the solid line. As you can see, this single line cannot capture many of the data points. On the right, we have an example of our GEEE expectile regression model (with an independent working correlation). As you can see, the five GEEE fitted models $(\tau \in (0.1, \ 0.25, \ 0.5, \ 0.75, \ 0.9))$ are able to capture many more data points in this heteroscedastic sample.


More specifically, the left and right panel of Figure \ref{fig:illustration_plot} display the same scatterplot. A classical mean regression is fitted to the data and represented by the fitted line in Figure \ref{fig:illustration_plot}.a. Although the mean regression parameter estimator is unbiased, the estimation of the location parameter alone does not capture the dispersion or the scale shifts caused by changes in the regressor, as suggested by the plot in Figure \ref{fig:illustration_plot}.a. In contrast, the ER model by estimating different conditional expectiles of the response distribution is able to capture the heteroscedasticity present in the data (Figure \ref{fig:illustration_plot}.b). Thus, similar to the ER model, the GEEE model by estimating the conditional expectiles of the response distribution will capture the location and scale shift and other unobserved heterogeneity of the data.

In section \ref{Model_geee}, we define the expectile and introduce the ER for the cross-sectional data. We then introduce the Generalized Expectile Estimating Equations (GEEE) model and present an algorithm showing the extension of some common correlation structures. In section \ref{Asymptotic_geee}, we present the asymptotic properties of the GEEE estimator, as well as a robust estimator of its covariance matrix. In section \ref{Simulation_geee}, we conduct simulations to evaluate the estimator's performance on small samples. In section \ref{Application_geee}, we apply the GEEE model to a real data set (a clinical trial for a new labor pain medication) to uncover the unobserved heterogeneity of the data and capture the heterogeneity of the regressor effects. Section \ref{Conclusion_geee} concludes. Please note that theorem proofs are provided in the \textbf{supplementary material}. 


%% file: 3model.tex
\section{Models and Methods} \label{Model_geee}
This section introduces the univariate expectile and the $\ER$ model. 

\textbf{Notations.} Vectors are written in lower bold letters, $\boldsymbol{x}\in\mathbb{R}^p,$ and matrices are represented in capital bold letters, $\boldsymbol{X}\in\mathbb{R}^{n\times p}.$ Estimated quantities are represented with a hat symbol $\widehat{\boldsymbol{X}},$ the inverse of a squared matrix is noted as $\boldsymbol{X}^{-1}$ and the transposed matrix is $\boldsymbol{X}\transpose.$ The symbols $\mathbb{I}_{p\times p}$ and $\mathbb{I}_{p}$ represent the identity matrix and are noted without subscript $(\mathbb{I})$ when the dimension is implicitly known. The symbol $\mathds{1}(t<0)$ is the indicator function and is equal to 1 if $t<0$ and to 0 otherwise. The vectors $\boldsymbol{1}$ and $\boldsymbol{0}$ are constant vectors filled with 1s and 0s, respectively. The function $\norm{ \cdot }_{\infty}$ is the $\max$ norm and the $l_2$ norm is noted as $\norm{ \cdot }_2\ $ or $\ \norm{ \cdot }.$ The symbol $\otimes$ represents the Kronecker product.

\subsection{Expectile and expectile regression}
The expectile of a continuous random variable $Y$ is defined as the solution $\mu_{\tau}(Y)$ that minimizes the loss function 

\begin{equation}\label{exp1_geee}
\E\{\rho_\tau(Y-\theta)\}   
\end{equation}

over $\theta\in\mathbb{R}$ for a fixed value of $\tau\in(0,1).$ The function $\rho_\tau(\cdot)$ of the form 

\begin{equation*}
  \rho_\tau(t)=\lvert \tau-\mathds{1}(t\leq 0)\rvert \cdot t^2  
\end{equation*}

is the asymmetric square loss function that assigns weights $\tau$ and $1-\tau$ to positive and negative deviations, respectively.

By equating the first derivative of (\ref{exp1_geee}) to zero, the expectile can also be defined as solution of

\begin{equation}\label{exp2_geee}
\mu_{\tau}(Y)=\mu_{\tau}=\mu -\frac{1-2\tau}{1-\tau}\E\big[ \{Y-\mu_{\tau}(Y)\} \mathds{1}\{Y>\mu_{\tau}(Y)\}\big],
\end{equation}

where $\mu = \mu_{0.5}(Y) = \E(Y).$ This definition, presented by \citet{newey1987}, shows that the expectile is determined by the tail expectations of the distribution of $Y.$ Interestingly, the expectile can also be defined as

\begin{equation*}
   \mu_{\tau}= \E\Bigg[ \frac{\psi_{\tau}(Y-\mu_{\tau})}{\E\big[\psi_{\tau}(Y-\mu_{\tau})\big]}Y\Bigg],
\end{equation*}

where $\psi_{\tau}(t)=\lvert \tau-\mathds{1}(t\leq 0)\rvert$ is the check function. This latter definition, which is much more meaningful in the context of regression, reveals that expectiles, like the mean, are weighted averages.

Given a random sample, $\lbrace(y_i)\rbrace_{i=1}^{n},$ the $\tau$-th empirical expectile  

\begin{equation*}
    \widehat{\mu}_{\tau}=\sum_{i=1}^{n}\frac{\psi_{\tau}(y_i-\widehat{\mu}_{\tau})}{\sum_{i=1}^{n}\psi_{\tau}
    (y_i-\widehat{\mu}_{\tau})}y_i
\end{equation*}

is the solution that minimizes the empirical loss function

\begin{equation}\label{exp3_geee}
\frac{1}{n}\sum_{i=1}^{n}\rho_\tau(y_i-\theta).
\end{equation}

\citet{newey1987} have shown that the expectile function has attractive properties. There is a one to one function between the expectile and the c.d.f. of the random variable. That is, the expectile function summarizes the c.d.f. of the random variable $Y,$ but differently than the quantile statistic. Indeed, the quantile is an order statistic while the expectile is a weighted mean. Moreover, the expectile is location and scale equivariant, that is, for $s>0 \mbox{ and } t\in\mathbb{R}, \; \mu_{\tau}(sY+t)=s\mu_{\tau}(Y)+t.$ More details about the expectile properties  and  the $\ER$ model are given by \citet{Efron1991}.

To introduce the $\ER$ method, consider the classical linear regression 

\begin{equation}\label{lineareg_geee}
y_i=\boldsymbol{x}_{i}\transpose\boldsymbol{\beta} + \varepsilon_{i},
\end{equation}

where $y_i$ is the scalar response, $\varepsilon_{i}$ the random error, $\ \boldsymbol{x}_{i}\in \mathbb{R}^p$ the vector of covariates and $\boldsymbol{\beta} \in \mathbb{R}^p$ the unknown parameter that needs to be estimated. Under this framework, \citet{newey1987} specified the $\ER$ model for a fixed $\tau \in (0,1)$ as:

\begin{equation}\label{explineareg_geee}
\mu_{\tau}(y_i|\boldsymbol{x}_{i})=\boldsymbol{x}_{i}\transpose\boldsymbol{\beta}_{\tau}, \quad \mu_{\tau}(\varepsilon_{i})=0.
\end{equation}

The assumption, $\mu_{\tau}(\varepsilon_{i})=0,$ ensures that the random error is centered on the $\tau$-th expectile. The corresponding $\ER$ estimator is defined as the unique solution that minimizes the objective function 

\begin{equation}\label{lr_est_geee}
\frac{1}{n}\sum_{i=1}^{n}\rho_\tau(y_i-\boldsymbol{x}_{i}\transpose\boldsymbol{\beta}_{\tau})
\end{equation}

over $\boldsymbol{\beta}_{\tau}\in\mathbb{R}^p.$ The parameter $\boldsymbol{\beta}_{\tau}$ measures the regressor effects and varies according to the value of $\tau$ in the presence of heteroscedasticity. In general, the choice of $\tau$ depends on the research question. For instance, if the interest is in studying the low birth weight risk factors, then the focus will be on the regressor effects at the extreme left of the response distribution, say $\tau \in (0.05, 0.1, 0.2).$
Regardless of the research question, \citet{newey1987} suggested a hypothesis testing procedure for detecting heteroscedasticity based on systematically estimating the regressor effects on a sequence of expectiles that spans the range of the response distribution and conducting multiple testing. This approach, in addition to testing for the presence of heteroscedasticity, will provide a comprehensive and detailed overview of the regressor effects on the response distribution.

The asymmetric loss function associated with the expectile function is continuously differentiable and solving equation model (\ref{lr_est_geee}) gives:

\begin{equation}\label{ler_est_geee}
\widehat{\boldsymbol{\beta}}_{\tau}=\Big(\sum_{i=1}^{n}\boldsymbol{x}_{i}\transpose\psi_{\tau}(\widehat{\varepsilon}_{i})\boldsymbol{x}_{i}\Big)^{-1}\Big(\sum_{i=1}^{n} \boldsymbol{x}_{i}\psi_{\tau}(\widehat{\varepsilon}_{i})y_{i}\Big),
\end{equation}

where $\widehat{\varepsilon}_{i}=y_{i}-\boldsymbol{x}_{i}\transpose\widehat{\boldsymbol{\beta}}_{\tau}.$ The $\ER$ estimator is easily computed using the iterated reweighted least squares (IRLS) algorithm. In addition to deriving the asymptotic properties of the above $\ER$ estimator, \citet{newey1987} proposed a robust estimator of the variance-covariance matrix for inference.   

Note that the $\ER$ estimator can be estimated through a likelihood-based approach. Indeed, assume that the disturbance, $u \sim \text{AND}(u;\mu_{\tau},\sigma^2,\tau),$ follows an asymmetric normal distribution (AND) 

\begin{equation}\label{and_geee}
f(u;\mu_{\tau},\sigma^2,\tau)=\frac{2}{\sqrt{\pi\sigma^2}}\frac{\sqrt{\tau(1-\tau)}}{\sqrt{\tau}+\sqrt{1-\tau}}
\exp\Bigg\lbrace-\rho_\tau\Bigg(\frac{u-\mu_{\tau}}{\sigma}\Bigg)\Bigg\rbrace,
\end{equation}

where $\mu_{\tau}, \ \sigma, \mbox{ and } \tau$ are the location, scale and asymmetric parameters, respectively. Now, substitute $\mu_{i\tau}=\boldsymbol{x}_{i}\transpose\boldsymbol{\beta}_{\tau}$ and assume that the $n$ observations are independent. Then, for a fixed $\tau,$ the $\ER$ estimator is equivalent to the maximum of the likelihood function 

\begin{equation*}
\mathbb{L}(\boldsymbol{\beta}; \sigma, \tau, \boldsymbol{y}) \propto \sigma^{-2n}
\exp\Bigg\lbrace-\sum_{i=1}^{n}\rho_\tau\Bigg(\frac{y_i-\boldsymbol{x}_{i}\transpose\boldsymbol{\beta}_{\tau}}{\sigma}\Bigg)
\Bigg\rbrace,
\end{equation*}

where $\boldsymbol{y}=(y_1,\ldots, y_n)\transpose.$ As mentioned by \citet{Waldmann2016}, the derived likelihood is technically not a likelihood but rather an auxiliary likelihood since it is not assumed to describe the exact or even approximately correct data distribution.

\subsection{GEEE for longitudinal data}

This section presents the model and method of the GEEE for longitudinal data analysis. Consider that the data $\lbrace y_{it},\boldsymbol{x}_{it}\rbrace_{1 \leq i \leq n, 1\leq t \leq m_i }$ are generated by the following model

\begin{equation}\label{m1_geee}
    y_{it} = \boldsymbol{x}_{it}\transpose\boldsymbol{\beta} + \varepsilon_{it}, 
\end{equation}

where $y_{it}$ is the $t$-th observation of the continuous response variable for the $i$-th individual, $\boldsymbol{x}_{it}=(x_{it}^1,\ldots,x_{it}^p)\transpose$ is the $p \times 1$ vector of regressors, $\varepsilon_{it}$ the random error and  $\boldsymbol{\beta}$ is the $p \times 1$ true parameter vector that needs to be estimated. By grouping observations from the same individual, equation model (\ref{m1_geee}) can be conveniently represented as 

\begin{equation}\label{m2_geee}
    \boldsymbol{y}_{i} = \boldsymbol{X}_{i}\boldsymbol{\beta} + \boldsymbol{\varepsilon}_{i}, 
\end{equation}

where $\boldsymbol{y}_{i}$ is the vector response of individual $i,$  $\ \boldsymbol{X}_{i}$ is the corresponding regressor matrix of dimension $m_i  \times p$ and $\boldsymbol{\varepsilon}_{i}$ the error vector. The individual observations can also be stacked and presented in matrix form as 

\begin{equation}\label{m3_geee} 
    \boldsymbol{y} = \boldsymbol{X}\boldsymbol{\beta} + \boldsymbol{\varepsilon},
\end{equation}

where $\boldsymbol{y}$ and $\boldsymbol{\varepsilon}$ are $N\times 1$ vectors, $\boldsymbol{X}$ is $N\times p$ matrix, and $N=\sum_{i=1}^n m_i.$

Using the location-scale equivariance property of the expectile function, we introduce the corresponding conditional $\tau$-expectile model as:

\begin{equation}\label{m4_geee}
    \mu_{\tau}(y_{it}|\boldsymbol{x}_{it})=\boldsymbol{x}_{it}\transpose\boldsymbol{\beta}_{\tau}, \quad \mu_{\tau}(\varepsilon_{it})=0.
\end{equation}

The assumption, $\mu_{\tau}(\varepsilon_{it})=0,$ is introduced to guarantee that the random error is centered on the $\tau$-th expectile. The parameter $\boldsymbol{\beta}_{\tau}$ measures the effect of the regressors on the expectile of the response variable distribution for a fixed $\tau\in(0,1).$ Thus, we can estimate the regressor effects on the response distribution by estimating the $\GEEE$ model for a sequence of expectiles $(\tau_1, \ldots, \tau_q).$ In this way, the $\GEEE$ captures the regressor effects on the location, scale and shape of the response distribution. 

A practical parameter estimator can be obtained by looking for the solution of the following expectile estimating equations:

\begin{equation}\label{indGEEE_geee}
\boldsymbol{S}_{I}(\boldsymbol{\beta}_{\tau})=\sum_{i=1}^{n} \frac{\partial\mu_{\tau}\Big(\boldsymbol{y_{i}|\boldsymbol{X}_i}\Big)}{\partial\boldsymbol{\beta}}\boldsymbol{\Psi}_{\tau}(\boldsymbol{y}_{i}-\boldsymbol{X}_i\boldsymbol{\beta}_{\tau})\Big[\boldsymbol{y}_{i}-
\mu_{\tau}(\boldsymbol{y_{i}|\boldsymbol{X}_i})\Big]= \boldsymbol{0},
\end{equation}

where $\boldsymbol{\Psi}_{\tau}(\boldsymbol{y}_{i}-\boldsymbol{X}_i\boldsymbol{\beta}_{\tau})=\diag\Big(\psi_{\tau}(y_{i1}-\boldsymbol{x}_{i1}\transpose\boldsymbol{\beta}_{\tau}),\ldots,\psi_{\tau}(y_{im_{i}}-\boldsymbol{x}_{im_{i}}\transpose\boldsymbol{\beta}_{\tau})\Big).$ The resulting estimator, $\widehat{\boldsymbol{\beta}}_{I\tau}, \ $ can also be derived as the minimizer of the following objective function:

\begin{equation}\label{m5_geee}
  \frac{1}{N}\sum_{i=1}^{n}\sum_{t=1}^{m_i } \rho_{\tau}\Big( y_{it}-\boldsymbol{x}_{it}\transpose\boldsymbol{\beta}_{\tau}\Big)
\end{equation}

over $\boldsymbol{\beta}_{\tau}\;\in\;\mathbb{R}^p.$ The explicit form of the resulting estimator $(\widehat{\boldsymbol{\beta}}_{I\tau})$ is similar to the classical $\ER$ estimator defined in equation (\ref{ler_est_geee}). When $\tau=0.5,$ the estimator $\widehat{\boldsymbol{\beta}}_{I\tau}$ corresponds to the $\GEEE$ estimator introduced by \citet{LiangZeger1986} with an independent working correlation structure. This fact is exploited to extend the $\GEE$ to the Generalized Expectile Estimating Equations $(\GEEE).$ 

The $\GEEE$ method  models the underlying correlation structure from the same subject by formally including a hypothesized structure to account for the within-subject correlation. For a fixed $\tau,$ the $\GEEE$ estimator 
$\widehat{\boldsymbol{\beta}}_{\tau}$ is derived by solving the following $\GEEE$ equations

\begin{equation}\label{GEEE1_geee}
\boldsymbol{S}(\boldsymbol{\beta}_{\tau})=\sum_{i=1}^{n} \frac{\partial\mu_{\tau}\Big(\boldsymbol{y_{i}|\boldsymbol{X}_i}\Big)}{\partial\boldsymbol{\beta}} \boldsymbol{V}_{i\tau}^{-1}\boldsymbol{\Psi}_{\tau} (\boldsymbol{y}_{i}-\boldsymbol{X}_i\boldsymbol{\beta}_{\tau})
\Big[\boldsymbol{y}_{i}-\mu_{\tau}(\boldsymbol{y_{i}|\boldsymbol{X}_i})\Big]= \boldsymbol{0},
\end{equation}

where $\boldsymbol{V}_{i\tau} \ $ is a working covariance matrix represented as

\begin{equation}\label{covmat_geee}
\boldsymbol{V}_{i\tau}=\sigma_{\tau}^2 \boldsymbol{A}_{i\tau}^{\frac{1}{2}}\boldsymbol{R}_{i}(\boldsymbol{\alpha}_{\tau})
\boldsymbol{A}_{i\tau}^{\frac{1}{2}},
\end{equation}

and $\sigma_{\tau}^2$ is the nuisance parameter. $\boldsymbol{A}_{i\tau}$ is the $m_i  \times m_i$ diagonal matrix with the variance function $\nu\Big(\mu_{\tau}(\boldsymbol{y_{i}|\boldsymbol{X}_i})\Big)$ as the diagonal elements, and $\boldsymbol{R}_{i}(\boldsymbol{\alpha}_{\tau})$ is the working correlation matrix. 

The working correlation matrix $\boldsymbol{R}_{i}(\boldsymbol{\alpha}_{\tau})$ describes the within-subject correlation pattern along the $K\times 1$ vector parameter $\boldsymbol{\alpha}_{\tau}.$ \citet{LiangZeger1986} proposed several types of working correlation structures (independent, exchangeable, autoregressive, unstructured, etc.) for the case when $\tau=0.5.$ These working correlations are adapted and extended to the GEEE approach. The extension of some of the most common and popular ones are presented below. 

The independent GEEE working correlation structure $(\Ind)$ is the simplest form of working correlation with the identity matrix and is the structure assumed by the expectile estimating equations model presented in equation (\ref{indGEEE_geee}). 
The exchangeable GEEE correlation structure $(\Exc)$ is a simple extension of the independence working correlation. It assumes a common correlation, $\rho_{ts\tau}=\alpha_{\tau}, \ \forall t \ne s,$ between any pair of measurements.
The autoregressive GEEE $(\AR)$ correlation defines the correlation of a pair of observations as a decreasing function of their distance in time, $\rho_{ts\tau}=\alpha_{\tau}^{\lvert t-s \rvert}.$ This structure assigns the highest correlation to adjacent pairs of observations and the lowest correlation to distant pairs. The unstructured GEEE correlation structure, as its name suggests, imposes no structure to the correlation matrix and defines the correlations of all pairs of measurements differently without any explicit pattern, $\rho_{ts\tau}=\alpha_{ts\tau} \ \forall t \ne s.$

All these types of working correlation are usually unknown and must be estimated. They are estimated in the iterative fitting process using the current value of the parameter vector. Indeed, the estimators can be computed as an iterated reweighted least squares estimators. The algorithm for the exchangeable GEEE working correlation structure is summarized by the stepwise procedure of Algorithm \ref{geee_algo}.

\begin{algorithm}[H]
\DontPrintSemicolon
\SetAlgoLined
\SetKwProg{Proc}{Procedure}{}{}

\KwIn{Let $\ \widetilde{\boldsymbol{\beta}}_{\tau}^{(0)} \gets \ \widehat{\boldsymbol{\beta}}_{I\tau}, \ $ 
  where $\ \widehat{\boldsymbol{\beta}}_{I\tau}\ $ is the estimator defined by equation (\ref{indGEEE_geee}).}
 \While{ $\norm{\widehat{\boldsymbol{\beta}}_{\tau}^{(r)}-\widehat{\boldsymbol{\beta}}_{\tau}^{(r-1)}}_{\infty} \leq \; \zeta \quad$ }{
   Given $ \ \widetilde{\boldsymbol{\beta}}_{\tau}^{(r-1)}$ at the $(r-1)$-th step, update: \;   
   \begin{enumerate}  
   \setlength\itemsep{1.3em}
    \item     
        $\widehat{\sigma}_{\tau}^{2(r)} \leftarrow  \frac{1}{N-p}\sum_{i=1}^{n}
    \sum_{t=1}^{m_i }\psi_{\tau}(\widehat{\varepsilon}_{it\tau})^2\widehat{\varepsilon}_{it\tau}^2$
  \item $\widehat{\alpha}_{\tau}^{(r)} \leftarrow  \frac{1}{(N_1-p)\widehat{\sigma}_{\tau}^{2(r)}}
    \sum_{i=1}^{n}\sum_{t<s}^{m_i }\psi_{\tau}(\widehat{\varepsilon}_{it\tau})
    \widehat{\varepsilon}_{it\tau}\psi_{\tau}(\widehat{\varepsilon}_{is\tau})
    \widehat{\varepsilon}_{is\tau}, \quad N_1 = \frac{1}{2}\sum_{i=1}^{n}m_i (m_i -1)$  
    \item $    \widehat{\boldsymbol{\beta}}_{\tau}^{(r)} \leftarrow \widehat{\boldsymbol{\beta}}_{\tau}^{(r-1)} + \Big[ \sum_{i=1}^{n}\boldsymbol{X}_i\transpose\boldsymbol{V}_{i\tau}^{-1}(\widehat{\alpha}_{\tau}^{(r-1)})\boldsymbol{\Psi}_{\tau}(\widehat{\boldsymbol{\beta}}_{\tau}^{(r-1)})\boldsymbol{X}_i\Big]^{-1}
    \boldsymbol{S}(\widehat{\boldsymbol{\alpha}}_{\tau}^{(r-1)},\widehat{\boldsymbol{\beta}}_{\tau}^{(r-1)})$    
\end{enumerate}    
}
\textbf{Return} $\widehat{\boldsymbol{\beta}}_{\tau}$ \;
\caption{GEEE algorithm}\label{geee_algo}
\end{algorithm}

The parameter $\zeta$ is the convergence tolerance and the default value in our code implementation is set to $10^{-7}. \ $ 
Algorithm \ref{geee_algo} is an IRLS algorithm and its convergence has been shown by \citet{burrus_iterative_1994}. Furthermore, \citet{jiang_iterative_2007} has shown, under mild conditions, the convergence of an iterative estimating equations (IEE) procedure, which includes Algorithm \ref{geee_algo}. Additionally, the IEE algorithm converges at an exponential rate and its estimator is consistent and asymptotically efficient \citep{jiang_iterative_2007}.

In practice, Algorithm \ref{geee_algo} is computationally efficient and usually the number of iterations required to achieve convergence is between 3 and 5. We have never encountered convergence issues in our simulation studies and real data analysis. Nevertheless, like the unrestricted GEE correlation structure, the estimated unrestricted GEEE correlation structure matrix is not guaranteed to be invertible and numeric problems may be encountered \citep{HardinJames2003}.

Notice that the convergence criterion defined as the maximum absolute relative change in the parameter estimators, at each iteration, is implemented in the first GEE software \citep{karimZeger1989} and in standard statistical software such as the geepack R package \citep{geepack1, geepack2, geepack3} or the GEE SAS procedure \citep{SASdoc}. We also tested the algorithm based on the relative change in the objective function. Both convergence criteria lead to the same results, but the algorithm based on the parameter estimates relative change is computationally faster.

Algorithm \ref{geee_algo} also applies to other types of working correlation; one can simply choose the appropriate estimator of the parameter $\boldsymbol{\alpha}$ which is either a scalar or a vector, depending on the type of correlation structure. For example, for a autoregressive GEEE  $(\AR)$ working correlation structure, the scalar parameter $\alpha_{\tau}$ is estimated by 

\begin{equation*}
  \widehat{\alpha}_{\tau}=  \frac{1}{(N_2-p)\widehat{\sigma}_{\tau}^{2}}\sum_{i=1}^{n}
  \sum_{t<m_i -1}\psi_{\tau}(\widehat{\varepsilon}_{it\tau})\widehat{\varepsilon}_{it\tau}\psi_{\tau}
  (\widehat{\varepsilon}_{i,t+1,\tau})
  \widehat{\varepsilon}_{i,t+1,\tau}, \ N_2=\sum_{i=1}^{n}(m_i -1).
\end{equation*}

For a unstructured GEEE working correlation structure, every element of the $m_i (m_i +1)/2$-vector parameter $\boldsymbol{\alpha}_{\tau}$ is estimated by 

\begin{equation*}
  \widehat{\alpha}_{ts\tau}=  \frac{1}{(N-p)\widehat{\sigma}_{\tau}^{2}}\sum_{i=1}^{n}\psi_{\tau}(\widehat{\varepsilon}_{it\tau})\widehat{\varepsilon}_{it\tau}\psi_{\tau}(\widehat{\varepsilon}_{is\tau})\widehat{\varepsilon}_{is\tau}.
\end{equation*}

Generalization to other GEEE working correlations is straightforward.

In Section \ref{Asymptotic_geee}, it is shown that the GEEE estimator $\widehat{\boldsymbol{\beta}}_{\tau}$ is consistent and asymptotically normally distributed. In addition, the simulation results in Section \ref{Simulation_geee} show that the GEEE method yields a consistent and highly efficient estimator even with a misspecification of the true covariance structure.

\subsection{GEEE for a sequence of expectiles}

A sequence of expectiles is often necessary, usually the mean and a few other expectiles above and below the mean, to describe the regressor effects on the response distribution. Additionally, the simultaneous estimation allows them to share strength among each other and to gain better estimation accuracy than individually estimated ones \citep{LiuWu2011}. For a fixed sequence of expectiles, $\boldsymbol{\tau}=(\tau_1, \ldots, \tau_q),$ the GEEE estimating functions are defined as 

\begin{equation}\label{GEEE2_geee}
  \begin{split}
      \boldsymbol{S}(\boldsymbol{\beta}_{\boldsymbol{\tau}}) &= \sum_{k=1}^{q}\boldsymbol{S}_{\tau_k}(\boldsymbol{\beta}_{\tau_k}) \\
       &= \sum_{i=1}^{n}(\boldsymbol{W}\otimes\boldsymbol{X}_i)\transpose \boldsymbol{V}_{i\boldsymbol{\tau}}^{-1}\boldsymbol{\Psi}_{\boldsymbol{\tau}}\Big(\mathds{1}_q\otimes\boldsymbol{y}_{i}-(\mathbb{I}_q\otimes\boldsymbol{X}_i)\boldsymbol{\beta}_{\boldsymbol{\tau}}\Big)\Big[\mathds{1}_q\otimes\boldsymbol{y}_{i}-(\mathbb{I}_q\otimes\boldsymbol{X}_i)\boldsymbol{\beta}_{\boldsymbol{\tau}}\Big],
  \end{split}
\end{equation}

where $\boldsymbol{S}_{\tau_k}$ is defined in (\ref{GEEE1_geee}), and $\boldsymbol{W}=[\diag(w_k)]_{k=1}^{q}$ is the $q\times q$ matrix of weights controlling the relative influence of the $q$ expectiles. $\boldsymbol{V}_{i\boldsymbol{\tau}}=[\diag(\boldsymbol{V}_{i\tau_k})]_{k=1}^{q}$ is a $q m_i \times q m_i$ block-diagonal working covariance matrix.  
For any fixed $\tau_k,$ the expression of the $m_i \times m_i$ matrix $\boldsymbol{V}_{i\tau_k}$ is function of the nuisance parameters $(\sigma_{\tau_k}, \boldsymbol{\alpha}_{\tau_k})$ and is given by equation (\ref{covmat_geee}). The diagonal matrix of check or influence functions, $\boldsymbol{\Psi}_{\boldsymbol{\tau}},$ is defined as follows:

\begin{equation*}
\boldsymbol{\Psi}_{\boldsymbol{\tau}}\Big(\mathds{1}_q\otimes\boldsymbol{y}_{i}-(\mathbb{I}_q\otimes\boldsymbol{X}_i)\boldsymbol{\beta}_{\boldsymbol{\tau}}\Big)=\diag\Big(\boldsymbol{\Psi}_{\tau_1}(\boldsymbol{y}_{i}-\boldsymbol{X}_i\boldsymbol{\beta}_{\tau_1}),\ldots, \boldsymbol{\Psi}_{\tau_q}(\boldsymbol{y}_{i}-\boldsymbol{X}_i\boldsymbol{\beta}_{\tau_q})\Big).
\end{equation*}

The parameter $\boldsymbol{\beta}_{\boldsymbol{\tau}} = (\boldsymbol{\beta}_{\tau_1}, \ldots, \boldsymbol{\beta}_{\tau_q})\transpose$ is obtained using the iterative reweighted least squares algorithm, as shown above, for a single expectile.

In general, the choice of the relative weight $\boldsymbol{W}$ is guided by the choice of the expectiles, which in turn depends on the research question. That being said, we suggest estimating the regressor effects at different expectiles of the response distribution and giving more weight to the expectiles of interest. \citet{koenker_quantile_2004} suggested to select the weights and the associated expectiles analogously to the choice of discretely weighted $L$-statistics. For example, using Tukey’s trimean we would assign weights $\boldsymbol{w} = (0.25, \ 0.5, \ 0.25)$ to the expectiles $\boldsymbol{\tau} = (0.25, \ 0.5, \ 0.75).$ 

In the next section, the asymptotic properties of the GEEE estimator are presented for a sequence of expectiles.

\section{Asymptotic properties}\label{Asymptotic_geee}

This section presents the asymptotic properties of the GEEE estimator for several fixed expectiles \(\boldsymbol{\tau}.\) In the first step, the asymptotic properties of the GEEE estimator \(\widehat{\boldsymbol{\beta}}_{I\boldsymbol{\tau}}\) with the independent working correlation structure are presented. Subsequently, the asymptotic properties of the GEEE estimator \(\widehat{\boldsymbol{\beta}}_{\boldsymbol{\tau}}\) with a general correlation structure are derived. The main reason for presenting the results of \(\widehat{\boldsymbol{\beta}}_{I\boldsymbol{\tau}}\) separately is that it is also the estimator of the expectile regression for a marginal model. In the following section, we assume that \(n\rightarrow\infty,\ \) and \(\ m=\max_{1\leq i \leq n}m_i\ \) is fixed. The proofs of the theorems are available in the \textbf{supplementary material}.

\subsection{Asymptotic properties for the independent GEEE} \label{independent_geee}

To begin, we assume that the following conditions are met.

\textbf{A1}. The data \(\lbrace (\boldsymbol{y}_i,\boldsymbol{X}_i)\rbrace_{i=1}^{n}\) are independent across \(i, \ \) and

\begin{equation*}
\begin{split}
{}& \Var\Big[\boldsymbol{\Psi}_{\boldsymbol{\tau}}(\boldsymbol{\varepsilon}_{i\boldsymbol{\tau}})\boldsymbol{\varepsilon}_{i\boldsymbol{\tau}}\Big]= \E\Big[\boldsymbol{\Psi}_{\boldsymbol{\tau}}(\boldsymbol{\varepsilon}_{i\boldsymbol{\tau}})\boldsymbol{\varepsilon}_{i\boldsymbol{\tau}}\boldsymbol{\varepsilon}_{i\boldsymbol{\tau}}\transpose
\boldsymbol{\Psi}_{\boldsymbol{\tau}}(\boldsymbol{\varepsilon}_{i\boldsymbol{\tau}})\Big]=\boldsymbol{\Sigma}_{i\boldsymbol{\tau}},\ \mbox{ where } \ \boldsymbol{\varepsilon}_{i\boldsymbol{\tau}}=\Big(\boldsymbol{\varepsilon}_{i\tau_1}\transpose,\ldots,\boldsymbol{\varepsilon}_{i\tau_q}\transpose\Big)\transpose \\
{}& \boldsymbol{\varepsilon}_{i\tau_k}=(\varepsilon_{i1\tau_k},\ldots,\varepsilon_{im_i\tau_k})\transpose, \quad \varepsilon_{it\tau_k}=y_{it}-\boldsymbol{x}_{it}\transpose\boldsymbol{\beta}_{\tau_k} \;
\mbox{ and } \; \boldsymbol{\Psi}_{\boldsymbol{\tau}}(\boldsymbol{\varepsilon}_{i\boldsymbol{\tau}})=
\Big[\diag(\boldsymbol{\Psi}_{\tau_k}(
\boldsymbol{\varepsilon}_{i{\tau_k}}))\Big]_{k=1}^{q}. \\
\end{split}
\end{equation*}

\textbf{A2}. The limiting forms of the following matrices are positive definite:

\begin{equation*}
\begin{split}
    {}& \boldsymbol{D}_{I1} (\boldsymbol{\tau})= \lim_{\substack{ \mathllap{n} \rightarrow \mathrlap{\infty} }} \quad N^{-1}
    \sum_{i=1}^{n}(\boldsymbol{W}\otimes\boldsymbol{X}_i)\transpose \E[\boldsymbol{\Psi}_{\boldsymbol{\tau}}(\boldsymbol{\varepsilon}_{i\boldsymbol{\tau}})](\mathbf{I}_q\otimes\boldsymbol{X}_i),\\
    {}& \\
    {}& \boldsymbol{D}_{I0} (\boldsymbol{\tau})= \lim_{\substack{ \mathllap{n} \rightarrow \mathrlap{\infty} }} \quad N^{-1}
    \sum_{i=1}^{n}(\boldsymbol{W}\otimes\boldsymbol{X}_i)\transpose\boldsymbol{\Sigma}_{i\boldsymbol{\tau}}(\boldsymbol{W}\otimes\boldsymbol{X}_i). \\
\end{split}
\end{equation*}

\textbf{A3}. There exists a positive constant $M$ such that \(\; \max_{\substack{ 1\leq i \leq n \\ 1 \leq t \leq m_i} }\norm{x_{it}} < M.\)

Assumptions \textbf{A1}-\textbf{A3} are standard assumptions for longitudinal models. Condition \textbf{A1} ensures independence across individuals, but permits a within-dependency between observations of the same individual, and allows heterogeneity across individuals. Condition \textbf{A2} is a standard full rank condition. Observe that when \(\tau =1/2, \ \) then \( \ \boldsymbol{\Sigma}_{i0.5}=1/4\Var[\boldsymbol{\varepsilon}_{i0.5}] \ \) becomes the variance of \(\boldsymbol{\varepsilon}_i\) up to a factor and

\begin{equation*}
\boldsymbol{D}_{I0}=1/4\lim_{n \rightarrow\infty}N^{-1}\sum_{i=1}^{n}\boldsymbol{X}_i\transpose\Var[\boldsymbol{\varepsilon}_{i0.5}]
\boldsymbol{X}_i.
\end{equation*}

Considering

\begin{equation*}
\boldsymbol{D}_{I1}=1/2\lim_{n \rightarrow\infty} N^{-1}\sum_{i=1}^{n}\boldsymbol{X}_i\transpose\boldsymbol{X}_i,
\end{equation*}

we see that this factor disappears in the expression of the variance. Therefore, when \(\tau =1/2, \ \) the condition \textbf{A2} is reduced to a condition on the matrices 
\begin{equation*}
N^{-1}\sum_{i=1}^{n}\boldsymbol{X}_i\transpose\Var[\boldsymbol{\varepsilon}_i]\boldsymbol{X}_i \quad \mbox{ and } \quad N^{-1}\sum_{i=1}^{n}\boldsymbol{X}_i\transpose\boldsymbol{X}_i .
\end{equation*}

Condition \textbf{A3} is important both for the convergence and for the Lindeberg condition. The following Theorem states the results of the asymptotic properties of the independent GEEE estimator \( (\widehat{\boldsymbol{\beta}}_{I\boldsymbol{\tau}}) \) assuming an independent working correlation structure.

\begin{theorem} \label{thm:theo1}
Assume that \(\widehat{\boldsymbol{\beta}}_{I\boldsymbol{\tau}}\) is the solution of the estimating function (\ref{indGEEE_geee}) and suppose the data are generated by model (\ref{m1_geee}) and that conditions \textbf{A1}-\textbf{A3} are satisfied. If
\(\E\lvert \psi_{\tau_k}(\varepsilon_{it\tau_{k}})\rvert^{4+\nu}<\Delta \mbox{ and }\E\lvert \varepsilon_{it\tau_{k}} \rvert^{4+\nu}<\Delta\)
for some \(\nu>0\) and \(\Delta>0,\) then for every fixed sequence of expectiles \(\boldsymbol{\tau}=(\tau_1,\ldots,\tau_q),\)
\begin{equation*}
 \sqrt{N}\big(\widehat{\boldsymbol{\beta}}_{I\boldsymbol{\tau}}-\boldsymbol{\beta}_{\boldsymbol{\tau}}\big)\xrightarrow{d} \mathcal{N}\bigg(\boldsymbol{0},    
  \boldsymbol{D}_{I1}^{-1}(\boldsymbol{\tau})\boldsymbol{D}_{I0}(\boldsymbol{\tau})\boldsymbol{D}_{I1}^{-1}(\boldsymbol{\tau})\bigg).
\end{equation*}
\end{theorem}

To use this new estimator \(\widehat{\boldsymbol{\beta}}_{I\boldsymbol{\tau}}\) to make inference, an estimator of its VC-matrix is needed and a robust one is presented in \textbf{Theorem \ref{thm:theo2}}. This will make it possible to construct large sample confidence intervals or conduct hypotheses testing. The estimator presented in \textbf{Theorem \ref{thm:theo2}} is a generalization of the robust VC estimator proposed by \citet{White1980} and used in, among other things, multilevel analysis \citep{LiangZeger1986}. This estimator inherits the same property that is, it accounts for the within-subject correlation and the heteroscedasticity between subjects. In summary, the proposed VC-matrix estimator is a commonly advocated covariance matrix estimator for longitudinal data. To state the theorem, we introduce the following matrices:

\begin{equation*}
\begin{split}
    {}& \widehat{\boldsymbol{D}}_{I1}(\boldsymbol{\tau})= N^{-1} \sum_{i=1}^{n}(\boldsymbol{W}\otimes\boldsymbol{X}_i)\transpose\boldsymbol{\Psi}_{\boldsymbol{\tau}}(\widehat{\boldsymbol{\varepsilon}}_{i\boldsymbol{\tau}})(\mathbf{I}_q\otimes\boldsymbol{X}_i),\\
    {}& \\
    {}& \widehat{\boldsymbol{D}}_{I0}(\boldsymbol{\tau})=N^{-1}\sum_{i=1}^{n}(\boldsymbol{W}\otimes\boldsymbol{X}_i)\transpose\widehat{\boldsymbol{\Sigma}}_{i\boldsymbol{\tau}} 
    (\boldsymbol{W}\otimes\boldsymbol{X}_i)\\
\end{split}
\end{equation*}

where \(\widehat{\boldsymbol{\Sigma}}_{i\boldsymbol{\tau}}=\boldsymbol{\Psi}_{\boldsymbol{\tau}}(\widehat{\boldsymbol{\varepsilon}}_{i\boldsymbol{\tau}})\widehat{\boldsymbol{\varepsilon}}_{i\boldsymbol{\tau}}\widehat{\boldsymbol{\varepsilon}}_{i\boldsymbol{\tau}}\transpose\boldsymbol{\Psi}_{\boldsymbol{\tau}}(\widehat{\boldsymbol{\varepsilon}}_{i\boldsymbol{\tau}})\) and \(\widehat{\boldsymbol{\varepsilon}}_{i\boldsymbol{\tau}}\) is obtained by replacing \(\boldsymbol{\beta}_{\boldsymbol{\tau}}\) with \(\widehat{\boldsymbol{\beta}}_{I\boldsymbol{\tau}}\) in the expression of \(\boldsymbol{\varepsilon}_{i\boldsymbol{\tau}}.\) Then, we have \textbf{Theorem \ref{thm:theo2}}.

\begin{theorem}\label{thm:theo2}
Suppose the data are generated by model (\ref{m1_geee}) and that conditions \textbf{A1}-\textbf{A3} are satisfied. If
\(\E\lvert \psi_{\tau_k}(\widehat{\varepsilon}_{it\tau_{k}})\rvert^{4+\nu}<\Delta \mbox{ and }\E\lvert \varepsilon_{it\tau_{k}} \rvert^{4+\nu}<\Delta\) for some \(\nu>0\) and \(\Delta>0,\) then for every fixed sequence of expectiles \(\boldsymbol{\tau}=(\tau_1,\ldots,\tau_q),\)

\begin{equation*}
  \widehat{\boldsymbol{D}}_{I1}^{-1}(\boldsymbol{\tau}) \widehat{\boldsymbol{D}}_{I0}(\boldsymbol{\tau}) \widehat{\boldsymbol{D}}_{I1}^{-1}(\boldsymbol{\tau})
  \xrightarrow{p} 
  \boldsymbol{D}_{I1}^{-1}(\boldsymbol{\tau}) \boldsymbol{D}_{I0}(\boldsymbol{\tau})\boldsymbol{D}_{I1}^{-1}(\boldsymbol{\tau}).
\end{equation*}
\end{theorem}

\subsection{Asymptotic properties for the general GEEE estimator} \label{general_geee}

After presenting the asymptotic properties of the GEEE-independent working correlation estimator, this subsection presents the asymptotic properties of the GEEE-estimator for a general working correlation. Assume that the following hold.

\textbf{B1}. The data \(\lbrace (\boldsymbol{y}_i,\boldsymbol{X}_i)\rbrace_{i=1}^{n}\) are independent across \(i\) and
\(\Var\Big[\boldsymbol{\Psi}_{\boldsymbol{\tau}}(\boldsymbol{\varepsilon}_{i\boldsymbol{\tau}})\boldsymbol{\varepsilon}_{i\boldsymbol{\tau}}\Big]= \boldsymbol{\Sigma}_{i\boldsymbol{\tau}}.\)

\textbf{B2}. The limiting forms of the following matrices are positive definite:

\begin{equation*}
\begin{split}
    {}& \boldsymbol{D}_{1} (\boldsymbol{\tau})= \lim_{\substack{ \mathllap{n} \rightarrow \mathrlap{\infty} }} \quad N^{-1}
    \sum_{i=1}^{n}(\boldsymbol{W}\otimes\boldsymbol{X}_i)\transpose\boldsymbol{V}_{i\boldsymbol{\tau}}^{-1}\E[\boldsymbol{\Psi}_{\boldsymbol{\tau}}(\boldsymbol{\varepsilon}_{i\boldsymbol{\tau}})](\mathbf{I}_q\otimes\boldsymbol{X}_i),\\
    {}& \\
    {}& \boldsymbol{D}_{0} (\boldsymbol{\tau})= \lim_{\substack{ \mathllap{n} \rightarrow \mathrlap{\infty} }} \quad N^{-1}
    \sum_{i=1}^{n}(\boldsymbol{W}\otimes\boldsymbol{X}_i)\transpose \boldsymbol{V}_{i\boldsymbol{\tau}}^{-1}\boldsymbol{\Sigma}_{i\boldsymbol{\tau}}\boldsymbol{V}_{i\boldsymbol{\tau}}^{-1}
    (\boldsymbol{W}\otimes\boldsymbol{X}_i).\\
\end{split}
\end{equation*}

\textbf{B3}. There exists a positive constant $M$ such that \(\max_{\substack{ 1\leq i \leq n \\ 1 \leq t \leq m_i} }\norm{x_{it}} < M.\)

The following theorem derives the asymptotic properties of the GEEE estimator with a general working correlation under the above conditions.

\begin{theorem}\label{thm:theo3}
Suppose the data are generated by model (\ref{m1_geee}) and that conditions \textbf{B1}-\textbf{B3} are satisfied. If
\(\E\lvert \psi_{\tau_k}(\varepsilon_{it\tau_{k}})\rvert^{4+\nu}<\Delta \mbox{ and }\E\lvert \varepsilon_{it\tau_{k}} \rvert^{4+\nu}<\Delta\)
for some \(\nu>0\) and \(\Delta>0,\) then for every fixed sequence of expectiles \(\boldsymbol{\tau}=(\tau_1,\ldots,\tau_q),\)

\begin{equation*}
 \sqrt{N}\big(\widehat{\boldsymbol{\beta}}_{\boldsymbol{\tau}}-\boldsymbol{\beta}_{\boldsymbol{\tau}}\big)\xrightarrow{d} \mathcal{N}\bigg(\boldsymbol{0}, \boldsymbol{D}_{1}^{-1}(\boldsymbol{\tau})
    \boldsymbol{D}_{0}(\boldsymbol{\tau})\boldsymbol{D}_{1}^{-1}(\boldsymbol{\tau})\bigg).
\end{equation*}
\end{theorem}

In the same way as with the GEEE-independent working correlation estimator, the next Theorem proposes an estimator of the VC-matrix of \(\widehat{\boldsymbol{\beta}}_{\boldsymbol{\tau}}.\) Consider

\begin{equation*}
\begin{split}
    {}& \widehat{\boldsymbol{D}}_{1}(\boldsymbol{\tau})= N^{-1}
    \sum_{i=1}^{n}(\boldsymbol{W}\otimes\boldsymbol{X}_i)\transpose\widehat{\boldsymbol{V}}_{i\boldsymbol{\tau}}^{-1}
\boldsymbol{\Psi}_{\boldsymbol{\tau}}(\widehat{\boldsymbol{\varepsilon}}_{i\boldsymbol{\tau}})(\mathbf{I}_q\otimes\boldsymbol{X}_i),\\
    {}& \\
    {}& \widehat{\boldsymbol{D}}_{0}(\boldsymbol{\tau})= N^{-1}
    \sum_{i=1}^{n}(\boldsymbol{W}\otimes\boldsymbol{X}_i)\transpose    \widehat{\boldsymbol{V}}_{i\boldsymbol{\tau}}^{-1}\widehat{\boldsymbol{\Sigma}}_{i\boldsymbol{\tau}}\widehat{\boldsymbol{V}}_{i\boldsymbol{\tau}}^{-1}
    (\boldsymbol{W}\otimes\boldsymbol{X}_i), \\
\end{split}
\end{equation*}
where $\widehat{\boldsymbol{\Sigma}}_{i\boldsymbol{\tau}}=\boldsymbol{\Psi}_{\boldsymbol{\tau}}(\widehat{\boldsymbol{\varepsilon}}_{i\boldsymbol{\tau}})\widehat{\boldsymbol{\varepsilon}}_{i\boldsymbol{\tau}}\widehat{\boldsymbol{\varepsilon}}_{i\boldsymbol{\tau}}
\transpose\boldsymbol{\Psi}_{\boldsymbol{\tau}}(\widehat{\boldsymbol{\varepsilon}}_{i\boldsymbol{\tau}}).$ The estimated working correlation $\widehat{\boldsymbol{V}}_{i\boldsymbol{\tau}} = \boldsymbol{V}_{i\boldsymbol{\tau}}(\widehat{\boldsymbol{\alpha}})$ is assumed to be a consistent estimator of $\boldsymbol{V}_{i\boldsymbol{\tau}}.$ This assumption is equivalent to assuming the existence of a consistent estimator, like the moment estimator, of the parameter $\boldsymbol{\alpha}.$ In the following theorem, we state the consistency result of our robust covariance estimator.

\begin{theorem}\label{thm:theo4}
Suppose the data are generated by model (\ref{m1_geee}) and that conditions \textbf{B1}-\textbf{B3} are satisfied. Assume
\(\E\lvert \psi_{\tau_k}(\widehat{\varepsilon}_{it\tau_{k}})\rvert^{4+\nu}<\Delta \mbox{ and }\E\lvert \varepsilon_{it\tau_{k}} \rvert^{4+\nu}<\Delta\) for some \(\nu>0\) and \(\Delta>0.\) Then for every fixed sequence of expectiles \(\boldsymbol{\tau}=(\tau_1,\ldots,\tau_q),\) and under the above conditions,

\begin{equation*}
  \widehat{\boldsymbol{D}}_{1}^{-1}(\boldsymbol{\tau}) \widehat{\boldsymbol{D}}_{0}(\boldsymbol{\tau}) \widehat{\boldsymbol{D}}_{1}^{-1}(\boldsymbol{\tau})
  \xrightarrow{p} 
  \boldsymbol{D}_1^{-1}(\boldsymbol{\tau}) \boldsymbol{D}_0(\boldsymbol{\tau})\boldsymbol{D}_1^{-1}(\boldsymbol{\tau}).
\end{equation*}
\end{theorem}

\textbf{Theorems} \ref{thm:theo3} and \ref{thm:theo4} allow for hypothesis testing about the population values of the parameter, $\boldsymbol{\beta}_{\boldsymbol{\tau}}.$ For instance, the presence of heterogeneous regressor effects can be detected by testing for differences in the vector of the slope coefficients across different expectiles.

Notice that by replacing the general correlation matrix with the identity matrix, the GEEE estimator proposed in Section \ref{general_geee} is the same as the independent GEEE estimator in section \ref{independent_geee}.

%% file: 4simulation.tex
\section{Simulations}\label{Simulation_geee}
\subsection{Model design}

In this section, the small sample performance of the GEEE estimators is evaluated through extensive simulation studies. The random samples are generated from the following linear model:

\begin{equation}\label{sim_mod_geee}
 M_{\gamma}: \quad   y_{it}=\beta_0 + x_{1it}\beta_1 + x_{2it}\beta_2 + (1 + \gamma x_{2it})\varepsilon_{it}, \; i=1,\ldots,n \ \mbox{ and } \ t = 1, \ldots, m_i.
\end{equation}

Two versions of equation model (\ref{sim_mod_geee}) are considered with respect to the parameter \(\gamma\in\lbrace 0, 3/10 \rbrace.\) A location-shift model \((M_{0})\) corresponding to \(\gamma=0,\) which assesses the performance of the estimators for an homoscedastic scenario; and a location-scale-shift model \((M_{3/10})\) corresponding to \(\gamma=3/10,\) serving to assess the performance of the estimators in the presence of heteroscedasticity.

The corresponding $\GEEE$ model for \((M_{0})\) is:  \(\ \mu_{\tau}(y_{it})=\beta_{0\tau} + x_{1it}\beta_1 + x_{2it}\beta_2,\ \) where 
\(\ \beta_{0\tau}=\beta_{0} +\mu_{\tau}(\varepsilon_{it}).\) In this model only the intercept varies with \(\tau\) and the expectile functions are parallel lines. In the location-scale-shift model \((M_{3/10}),\) the GEEE model is specified as: \(\ \mu_{\tau}(y_{it})=\beta_{0\tau} + x_{1it}\beta_1 + x_{2it}\beta_{2\tau}, \ \)
where \(\ \beta_{0\tau}= \beta_{0} + \mu_{\tau}(\varepsilon_{it}\ ) \mbox{ and } \ \beta_{2\tau}= \beta_{2} + \gamma\mu_{\tau}(\varepsilon_{it}).\) Therefore, in the presence of heteroscedasticity, both the intercept and the parameter of $x_{2}$ will vary according to \(\tau\), while the parameter of $x_{1}$ remains constant.

We generated a between-subject regressor \((x_1)\) from a binomial distribution $\mathcal{B}(1, 0.5)$ and a within-subject regressor $(x_2)$ from a Gaussian distribution. The parameters $\beta_0, \ \beta_1$ and $\ \beta_2$ were set to $(0.7, \ 0.4, \ 1.2),$ respectively. Finally, we generated the random error of equation model (\ref{sim_mod_geee}) from three distinct distributions: Normal $(\mathcal{N}(0,1)),$ Student with three degrees of freedom $(\mathcal{T}_3)$ and Chi-squared with three degrees of freedom $(\chi_2(3)).$ We used the R package copula \citep{copulaPack} to simulate the dependency between measurements of the same subject. We started by simulating a dependent uniform margins from a Gaussian copula with an $\AR$ correlation structure. Then, we generated the dependent random errors as quantiles of the uniform margins from the three distinct marginal distributions $(\mathcal{N}(0,1), \ \mathcal{T}_3, \ \chi_2(3)).$ Specifically, we generated the data as follows:

\begin{enumerate}
    \item Generate the predictors $x_1$ and $x_2;$
    \item Generate a uniform sample: $(u_1,\ldots, u_{m_{i}})$ from a Gaussian Copula with an $\AR$ correlation structure and a specific correlation parameter $\rho;$ 
    \item For $t=1,\ldots,m_i$, \ generate the dependent random error  $\varepsilon_{it}^{\prime}=F^{-1}(u_{it}),$
    where $F(.)$ is one of the three marginal distributions $(\mathcal{N}(0,1), \ \mathcal{T}_3, \ \chi_2(3));$ and 
    \item Generate the response variable: $y_{it}=\beta_0 + x_{1it}\beta_1 + x_{2it}\beta_2 + (1 + \gamma x_{2it})\varepsilon_{it}.$
\end{enumerate}

We used three different values for the correlation parameter: low \((\rho=0.1),\) medium \((\rho=0.5),\) and high \((\rho=0.9)\) correlations. We generated each model according to three different sample sizes \(n\in\lbrace 50, \ 100, \ 250 \rbrace.\) Finally, for the number of repeated measurements \(m_i,\) a balanced design with \(m_i=4\) and an unbalanced design were studied. In the unbalanced design, \(m_i\) is a number randomly generated between 3 to 7 with equal probability. The extensive simulation was carried out with 400 replications for each parameter-combination scenario. In each scenario, the focus was on the effect of the regressors at the expectiles \(\tau\in\lbrace 0.1,\ 0.2,\ 0.3,\ 0.4,\ 0.5,\ 0.6,\ 0.7,\ 0.8,\ 0.9 \rbrace.\) All computations were performed with the \texttt{R} (v4.0.1) statistical programming language \citep{RCoreTeam2018}. The implemented R package \textbf{expectgee} that comes with this manuscript is publicly available on GitHub at \url{github.com/AmBarry/expectgee}.

\subsection{Performance measures}
We fitted the simulated samples using the GEEE model with three working correlation structures: independence (Ind), exchangeable $(\Exc)$ and $\AR$ correlations, and studied their performance in terms of bias, relative efficiency and model selection. We reported the  bias distribution  of the estimators as an error plot (the mean and the range of the distribution). We evaluated the relative efficiency $(\Reff)$  by the ratio of the standard errors and by taking the GEEE standard error with an $\AR$ correlation structure (which is the true correlation structure) as denominator. The relative efficiency $(\Reff)$ of model $M \in (\Ind, \ \Exc, \ \AR)$ was defined as:

\begin{equation*}
\Reff_{M} = \frac{SE_{M}(\beta_{k\tau})}{SE_{\AR}(\beta_{k\tau})}, \quad k \in \lbrace 0, 1, 2\rbrace. 
\end{equation*}
where $SE_{M}(\beta_{k\tau}) = \frac{1}{400} \sum_{j=1}^{400} \widehat{\sigma}^{(j)}_{\beta_{k\tau}}$ and $\widehat{\sigma}_{\beta_{k\tau}}$ is the estimated standard error of $\beta_{k\tau}.$

We also evaluated the performance of the asymptotic standard error (SE) presented in Theorem \ref{thm:theo4} by reporting the distribution of the ratio between the asymptotic standard error (SE) and the Monte Carlo standard deviation (SD) defined as: 

\begin{equation*}
\text{SD}_{M}^{2}(\beta_{k\tau}) = \frac{1}{400}  \sum_{j=1}^{400} \Big(\widehat{\beta}^{(j)}_{k\tau} - \overline{\widehat{\beta}}_{k\tau}\Big)^2 \, \quad k \in \lbrace 0, 1, 2\rbrace, 
\end{equation*}

where $\overline{\widehat{\beta}}_{k\tau} = \frac{1}{400}  \sum_{j=1}^{400} \widehat{\beta}^{(j)}_{k\tau}.$ 

Since the efficiency of the $\GEEE$ parameter estimates depended on the accuracy of the correlation structure, it was important to select the most accurate correlation structure. Numerous criteria to select the best working correlation structure have been proposed and there was no criterion that seems to work better than others. Therefore, in this manuscript we relied on two popular selection criteria: the quasilikelihood under the independence model information criterion $(\QIC)$ and the correlation information criterion $(\CIC)$ to select the best working correlation  structure and to evaluate the goodness-of-fit of our GEEE model. The $\QIC$ criterion developed by \citet{Pan2001} is an adaptation of the Akaike Information Criterion for the GEE model and is defined as:

\begin{equation*}
\QIC(\boldsymbol{R}) = -2 Q(\widehat{\boldsymbol{\beta}}(\boldsymbol{R}), I, \Delta) + \Tra\Big( \widehat{\boldsymbol{\Omega}}_{\mathbf{I}}(\widehat{\boldsymbol{\beta}}(\boldsymbol{R})) \widehat{\boldsymbol{V}}(\widehat{\boldsymbol{\beta}}(\boldsymbol{R}))\Big),
\end{equation*}

where $Q$ is the quasilikelihood function \citep{Wedderburn1974} under independence assumption of the dataset $\Delta.$ In our case $Q=-1/2\sum_{it}\widehat{\varepsilon}^2_{it}.$ The matrix \(\widehat{\boldsymbol{\Omega}}_{\mathbf{I}}\) was the inverse of the variance matrix obtained by fitting an independence model and $\widehat{\boldsymbol{V}}$ was the sandwich variance estimator under the working correlation structure presented in \textbf{Theorem} \ref{thm:theo4}. \citet{cicCriterion2009} argued that the first term in the $\QIC$ criterion which is the quasilikelihood under the independence assumption was not informative in the selection of the covariance structure and proposed using the second term alone for the selection of the working correlation structure. They called it $\CIC$ and defined it as:

\begin{equation*}
    \CIC(\boldsymbol{R})=\Tra\Big( \widehat{\boldsymbol{\Omega}}_{\mathbf{I}}(\widehat{\boldsymbol{\beta}}(\boldsymbol{R})) \widehat{\boldsymbol{V}}(\widehat{\boldsymbol{\beta}}(\boldsymbol{R}))\Big).
\end{equation*}

\citet{geePack2005} reported that the $\CIC$ criterion was more convenient compared to the $\QIC$ criterion, particularly when the model for the mean did not fit well the data and when the models with different correlation structures were compared.

We also defined an asymmetric version of these criteria and took their sum as a new criterion. The corresponding asymmetric version of the $\QIC$ and the $\CIC$ were defined as:

\begin{equation*}
   \asymQIC(\boldsymbol{R}) = \sum_{k=1}^{q} \QIC_{\tau_{k}}(\boldsymbol{R}_{\tau_{k}}) \; \mbox{ and } \;
   \asymCIC(\boldsymbol{R}) = \sum_{k=1}^{q} \CIC_{\tau_{k}}(\boldsymbol{R}_{\tau_{k}})
\end{equation*}

where $q$ is the number of selected expectiles. In this manuscript we selected 9 expectiles for the simulation and the real data application. The asymmetric measure $\QIC_{\tau}$ and $\CIC_{\tau}$ defined at $\tau\in (0,1)$ correspond to the classical $\QIC$ and $\CIC$ computed with the parameter estimate $(\widehat{\boldsymbol{\beta}}_{\tau})$ and the working correlation $\boldsymbol{R}_{\tau}$ of the asymmetric model, respectively. Notice that when $\tau=0.5,$ the asymmetric measures $\QIC_{\tau}$ and $\CIC_{\tau}$ corresponded to the classical $\QIC$ and $\CIC,$ respectively.

The asymmetric criteria $\asymQIC$ and $\asymCIC$ selected in majority the true correlation structure $(\AR).$ Furthermore, the $\asymQIC$ and $\asymCIC$ measures performed better than the $\QIC$ and $\CIC$ measures in the presence of heteroscedasticity, and when the error distribution was heavy-tailed, see the \textbf{supplementary material}. 

\begin{center}
\begin{figure}[hbt!]
\includegraphics[width=1\linewidth]{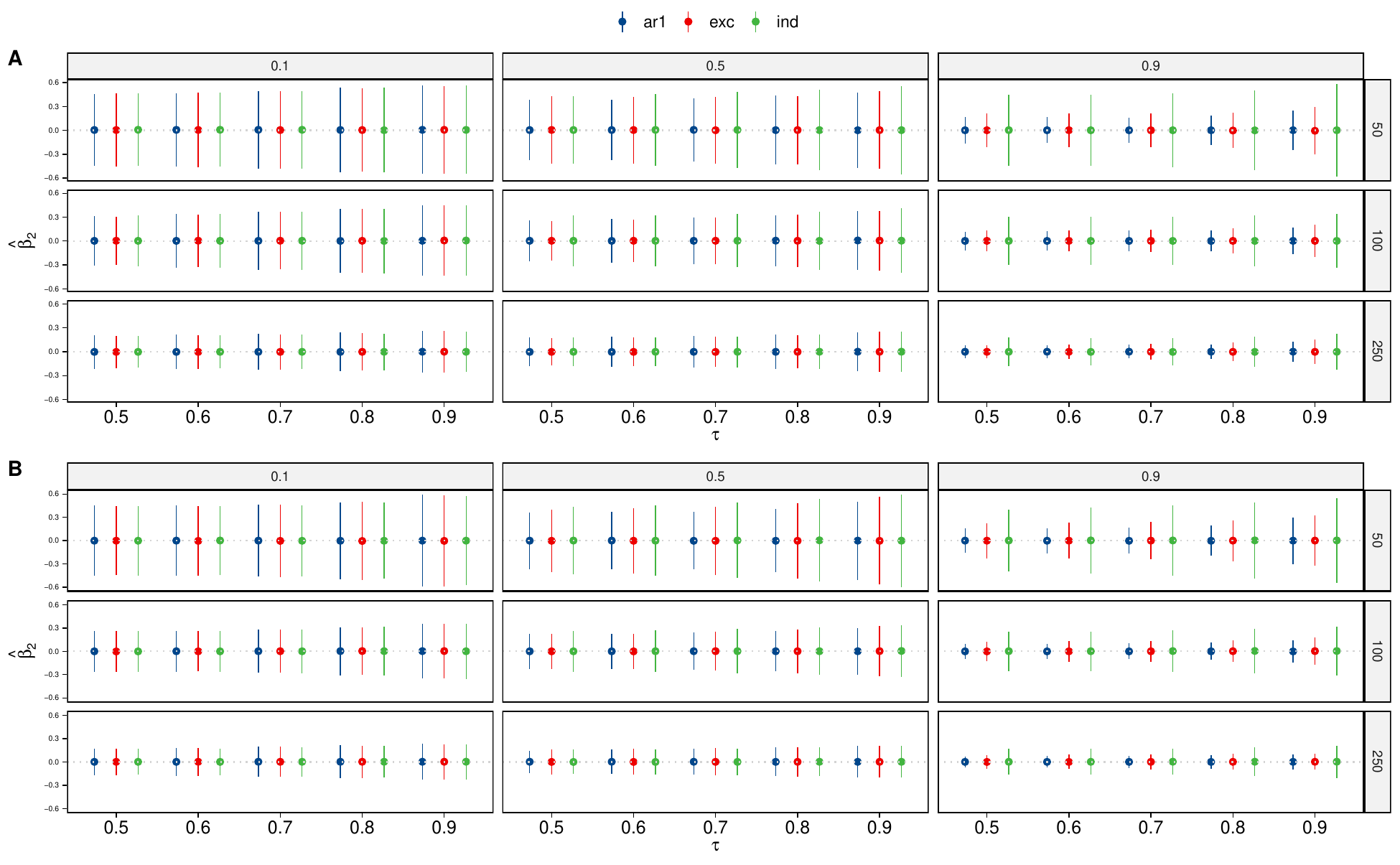}
  \caption{Bias distribution of $\widehat{\beta}_2$ represented as an error plot according to the sample size 
$n\in(50,  \ 100,  \ 250),$ the degree of correlation $\rho \in (0.1, \ 0.5, \ 0.9),$ the expectiles $\tau\in (0.5,  \ 0.6,  \  0.7, \  0.9)$
and the error term $\varepsilon\sim\mathcal{N}(0, \ 1)$ in a location-shift scenario. Figures \textbf{A}-\textbf{B} represent the results for the balanced $(m=4)$ and unbalanced panel $(m\sim\mathcal{U}(3, \ 7)),$ respectively.}\label{fig:b2bias_norm_homo}
\end{figure}
\end{center}

\subsection{Results}
We presented in this section the performance of the GEEE model in comparison to the GEE according to the previous defined performance measures. For the sake of space, we could not present all the simulation results in this document. We had chosen to present here the results of the parameter estimate of $x_2$ (which is correlated to the random error in the $M_{3/10}$ model) for the expectiles $\tau\in\lbrace 0.5,\ 0.6,\ 0.7,\ 0.8,\ 0.9 \rbrace$ and for a Gaussian random error (location and location-scale-shift). The simulation results of the other parameter estimates, distributions (Student and Chi-Squared) and expectiles $\tau\in\lbrace 0.1,\ 0.2,\ 0.3,\ 0.4,\ 0.5 \rbrace$ were postponed in the \textbf{supplementary material}.

\begin{center}
\begin{figure}[hbt!]
\includegraphics[width=1\linewidth]{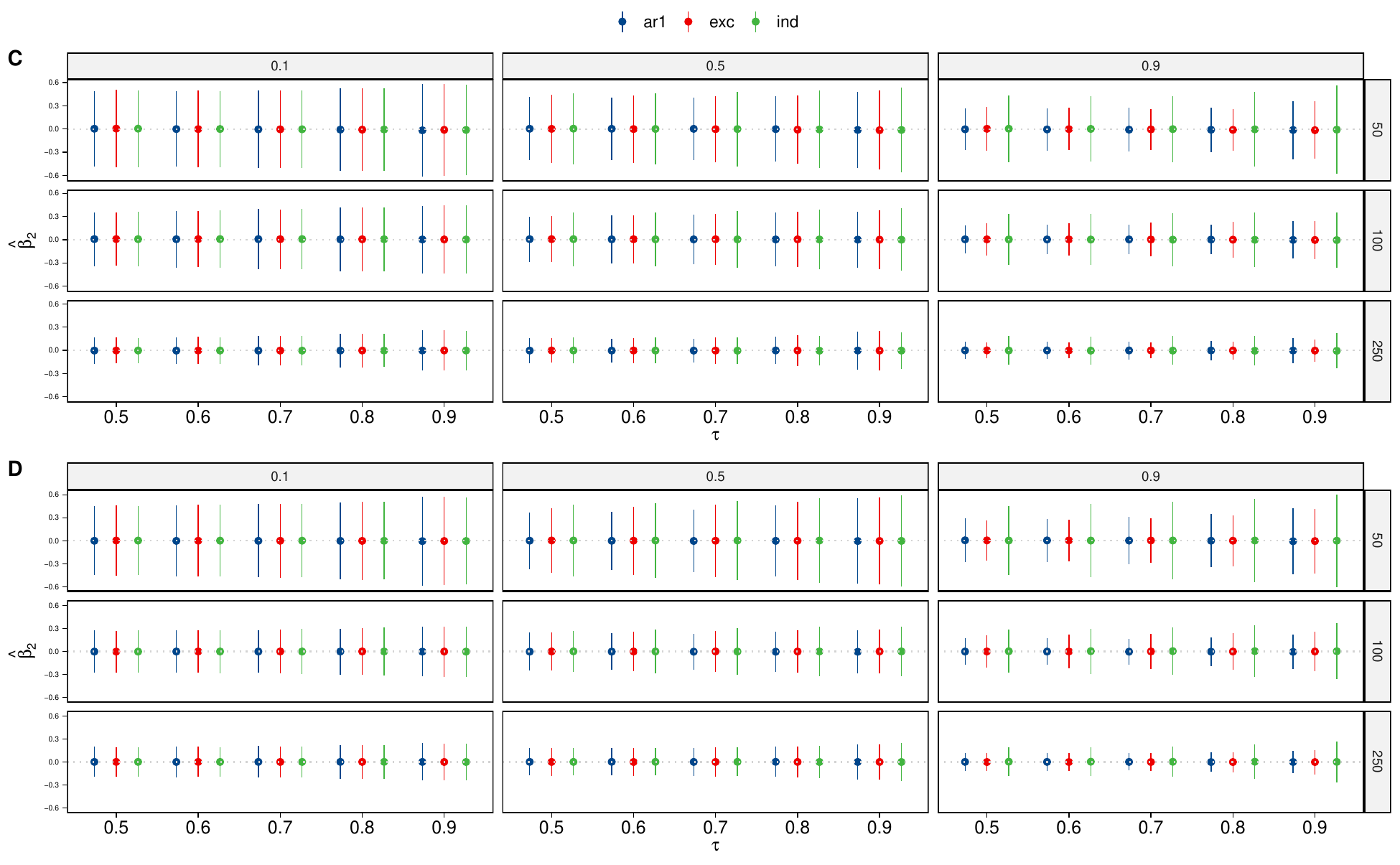}
  \caption{Bias distribution of $\widehat{\beta}_2$ represented as an error plot according to the sample size 
$n\in(50,  \ 100,  \ 250),$ the degree of correlation $\rho \in (0.1, \ 0.5, \ 0.9),$ the expectiles $\tau\in (0.5,  \ 0.6,  \  0.7, \  0.9)$
and the error term $\varepsilon\sim\mathcal{N}(0, \ 1)$ in a location-scale-shift scenario. Figures \textbf{C}-\textbf{D} represent the results for the balanced $(m=4)$ and unbalanced panel $(m\sim\mathcal{U}(3, \ 7)),$ respectively.}\label{fig:b2bias_norm_hetero}
\end{figure}
\end{center}

\textbf{Figure \ref{fig:b2bias_norm_homo}} and \textbf{Figure \ref{fig:b2bias_norm_hetero}} reported the bias distribution  in the form of an error plot centered on the mean, in the location-shift $(M_{0})$ and the location-scale-shift $(M_{3/10})$ scenarios, respectively. Overall, the bias was centered around 0 with a relatively small range as the sample size increased in the location-shift and location-scale-shift scenarios. The GEEE model with the $\AR$ working correlation structure{\textemdash} which is the true correlation structure, outperformed the GEEE model with the other working correlation structure $(\Exc, \Ind).$ Moreover, the bias distribution  of the $\GEEE$ models were similar to the bias distribution  of the classical $\GEE.$ The bias distribution  of the independent GEEE model displayed a larger range when the degree of correlation was high $(\rho=0.9),$ which suggested that a misspecification of the random error's correlation structure could introduce some bias. Similar performances were observed when the error was generated by a Student or a Chi-Squared distribution. These results can be found in the \textbf{supplementary material}.

\begin{center}
\begin{figure}[hbt!]
\includegraphics[width=1\linewidth]{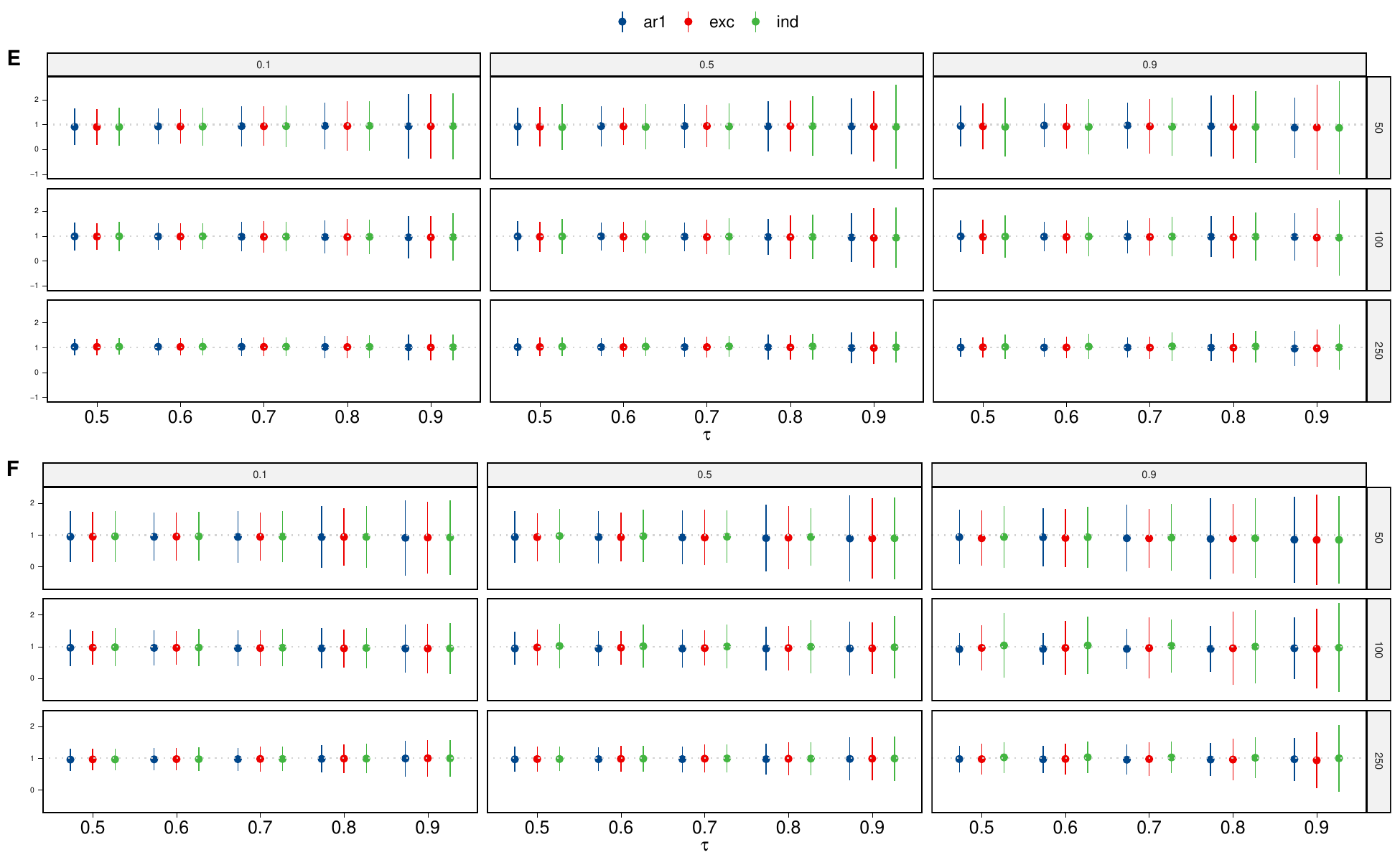}
  \caption{Distribution of the ratio $\frac{\SE}{\SD}$ for $\widehat{\beta}_2$ estimator represented as an error plot with respect to the sample size $n\in(50,  \ 100,  \ 250),$ the degree of correlation $\rho \in (0.1, \ 0.5, \ 0.9),$ the expectiles 
  $\tau\in (0.5,  \ 0.6,  \  0.7, \  0.9)$ and the error term $\varepsilon\sim\mathcal{N}(0, \ 1)$ in a location-shift scenario. Figures \textbf{E}-\textbf{F} represent the results for the balanced $(m=4)$ and unbalanced panel $(m\sim\mathcal{U}(3, \ 7)),$ respectively.} \label{fig:b2SdSe_norm_homo}
\end{figure}
\end{center}

\begin{center}
\begin{figure}[hbt!]
\includegraphics[width=1\linewidth]{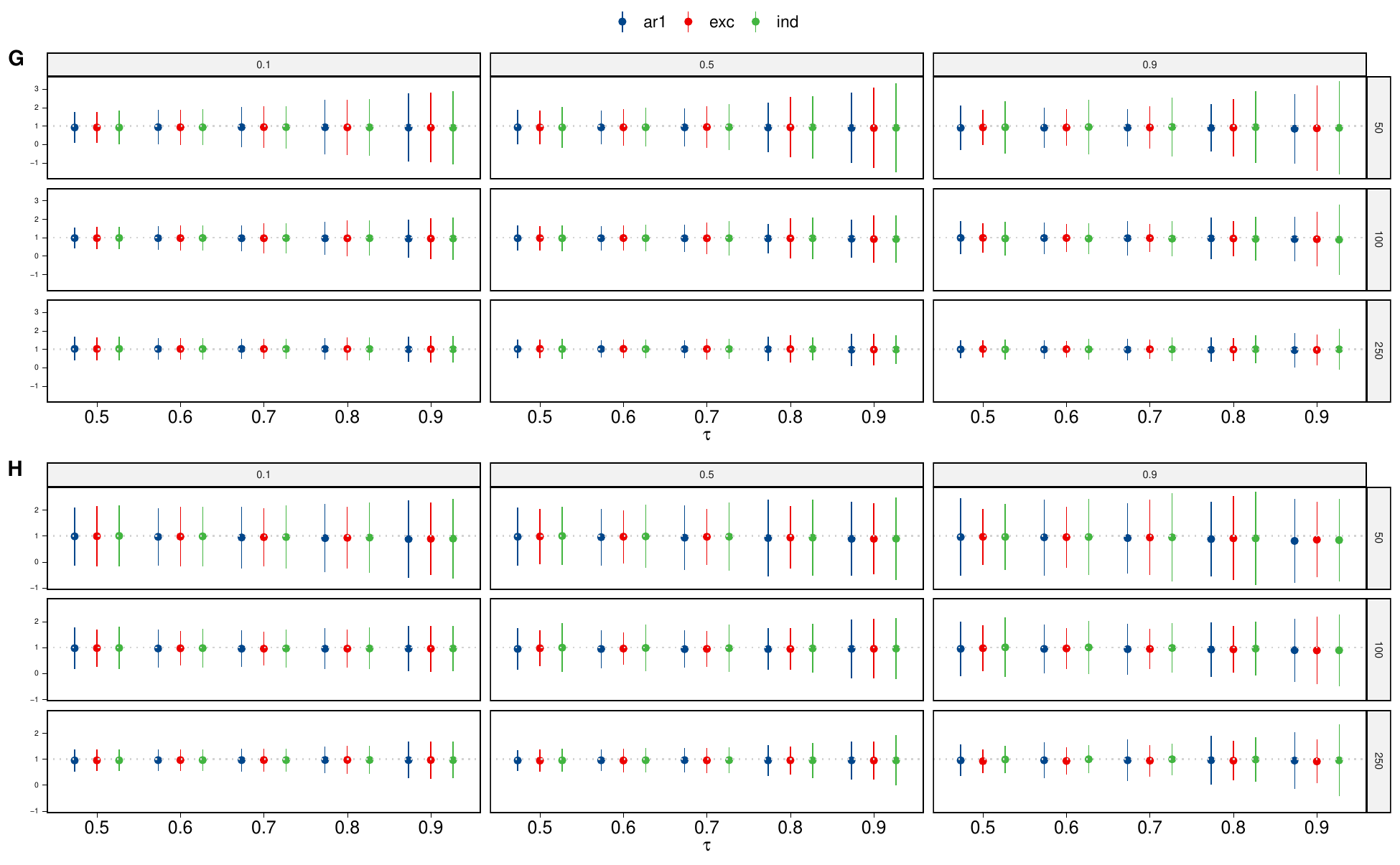}
  \caption{Distribution of the ratio $\frac{\SE}{\SD}$ for $\widehat{\beta}_2$ estimator represented as an error plot with respect to the sample size $n\in(50,  \ 100,  \ 250),$ the degree of correlation $\rho \in (0.1, \ 0.5, \ 0.9),$ the expectiles 
  $\tau\in (0.5,  \ 0.6,  \  0.7, \  0.9)$
and the error term $\varepsilon\sim\mathcal{N}(0, \ 1)$ in a location-scale-shift scenario. Figures \textbf{G}-\textbf{H} represent the results for the balanced $(m=4)$ and unbalanced panel $(m\sim\mathcal{U}(3, \ 7)),$ respectively.} \label{fig:b2SdSe_norm_hetero}
\end{figure}
\end{center}

To evaluate the asymptotic standard error $(\SE)$ of the $\GEEE$ parameter estimates, we used the Monte Carlo standard deviation $(\SD)$ as a benchmark and presented the distribution of the ratio $\frac{\SE}{\SD}$ as an error plot centered at the mean, \textbf{Figure \ref{fig:b2SdSe_norm_homo}}-\textbf{Figure \ref{fig:b2SdSe_norm_hetero}}. The results showed that the error plots were centered at 1, which means that on average the asymptotic standard error $\SE$ and the Monte Carlo standard deviation $\SD$ were identical. However, we observed a larger range of this ratio at the extreme expectiles, especially when the sample size was small. Furthermore, this variability was more pronounced for the $\Exc$ and $\Ind$ correlation structures than the $\AR$ correlation structure, especially when the degree of correlation was high $(\rho=0.9).$ Similar performances were observed for the Student and Chi-Squared random error. Those results can be found in the \textbf{supplementary material}.

\begin{center}
\begin{figure}[hbt!]
\includegraphics[width=1\linewidth]{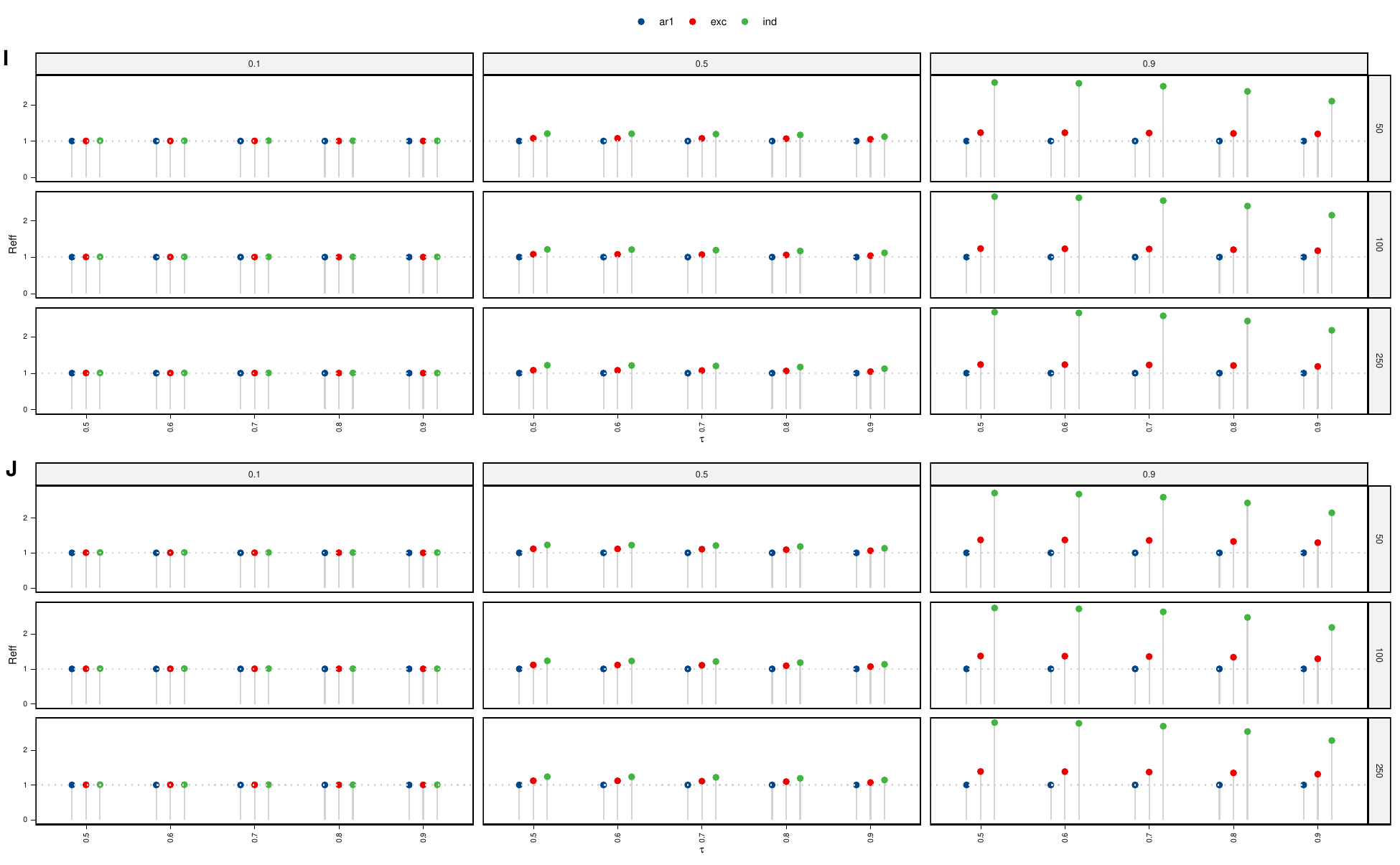}
  \caption{Relative efficiency $(\Reff)$ of the working correlation structures $(\AR, \Exc, \Ind)$ for $\widehat{\beta}_2$
estimator with respect to the sample size $n\in(50,  \ 100,  \ 250),$ the degree of correlation $\rho \in (0.1, \ 0.5, \ 0.9),$ the expectiles 
$\tau\in (0.5,  \ 0.6,  \  0.7, \  0.9)$ and the error term $\varepsilon\sim\mathcal{N}(0, \ 1)$ in a location-shift scenario. Figures \textbf{I}-\textbf{J} represent the results for the balanced $(m=4)$ and unbalanced panel $(m\sim\mathcal{U}(3, \ 7)),$ respectively.} \label{fig:b2Reff_norm_homo}
\end{figure}
\end{center}

\begin{center}
\begin{figure}[hbt!]
\includegraphics[width=1\linewidth]{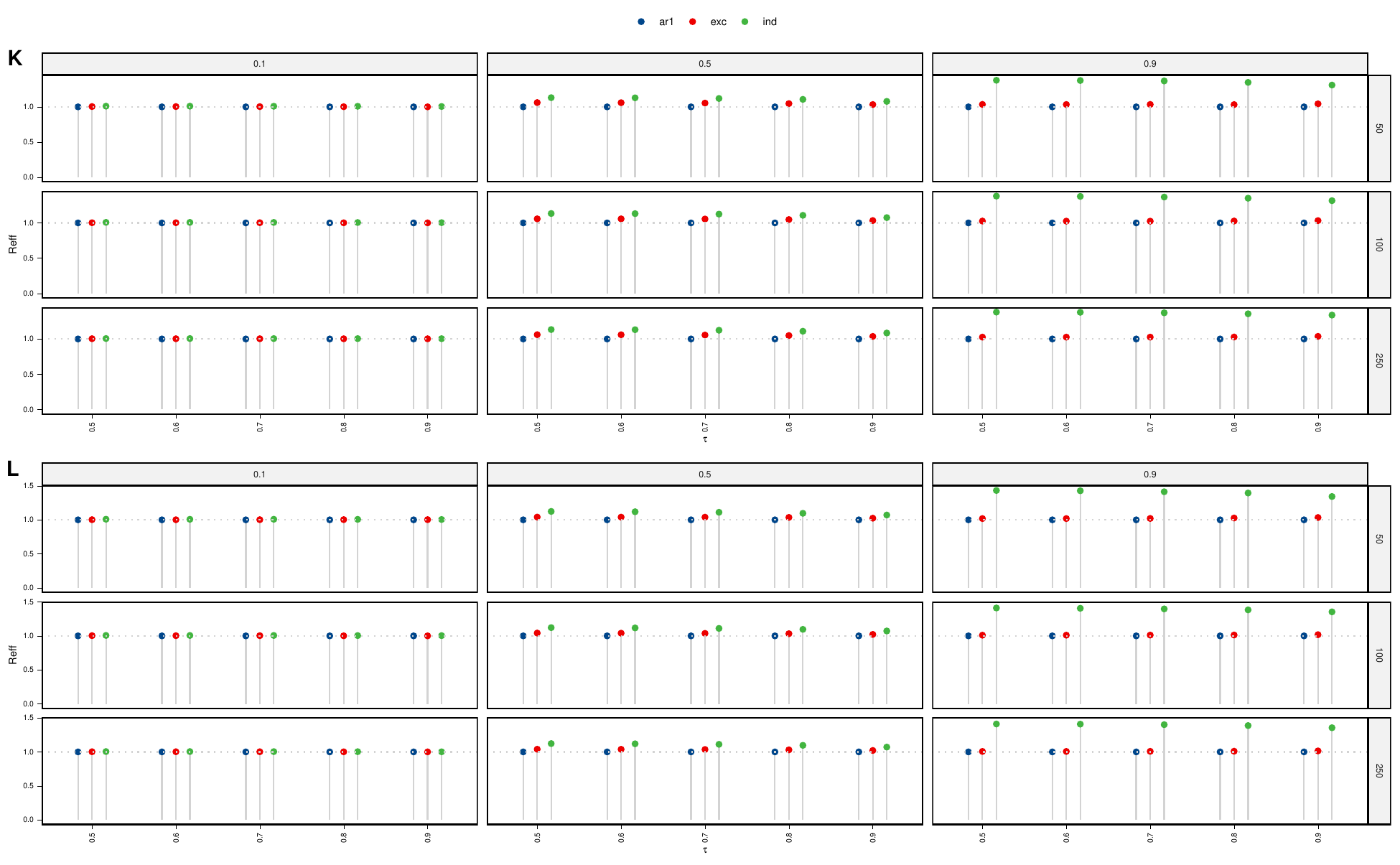}
  \caption{Relative efficiency $(\Reff)$ of the working correlation structures $(\AR, \Exc, \Ind)$ for $\widehat{\beta}_2$
estimator with respect to the sample size $n\in(50,  \ 100,  \ 250),$ the degree of correlation $\rho \in (0.1, \ 0.5, \ 0.9),$ the expectiles 
$\tau\in (0.5,  \ 0.6,  \  0.7, \  0.9)$ and the error term $\varepsilon\sim\mathcal{N}(0, \ 1)$ in a location-scale-shift scenario. Figures \textbf{K}-\textbf{L} represent the results for the balanced $(m=4)$ and unbalanced panel $(m\sim\mathcal{U}(3, \ 7)),$ respectively.}\label{fig:b2Reff_norm_hetero}
\end{figure}
\end{center}

We reported the results of the relative efficiency $(\Reff)$ of the different working correlation structures in \textbf{Figure \ref{fig:b2Reff_norm_homo}} and \textbf{Figure \ref{fig:b2Reff_norm_hetero}}, for the location-shift $(M_{0})$ and the location-scale-shift scenarios $(M_{3/10}),$ respectively. The $\Reff$ of the $\AR$ working correlation structure was equal to 1, because it was set as the reference model. We observed that the $\Exc$ and the $\Ind$ working correlation structures always had a relative efficiency greater than 1, $(\Reff>= 1).$ That is, the $\GEEE$ model with the $\AR$ working correlation structure{\textemdash} which is the true correlation structure, was more efficient than the other working correlation structures. We also noticed that the relative efficiency of the $\Ind$ working correlation structure was particularly higher when the degree of correlation was high $(\rho=0.9).$ 

\begin{center}
\begin{figure}[hbt!]
\includegraphics[width=1\linewidth]{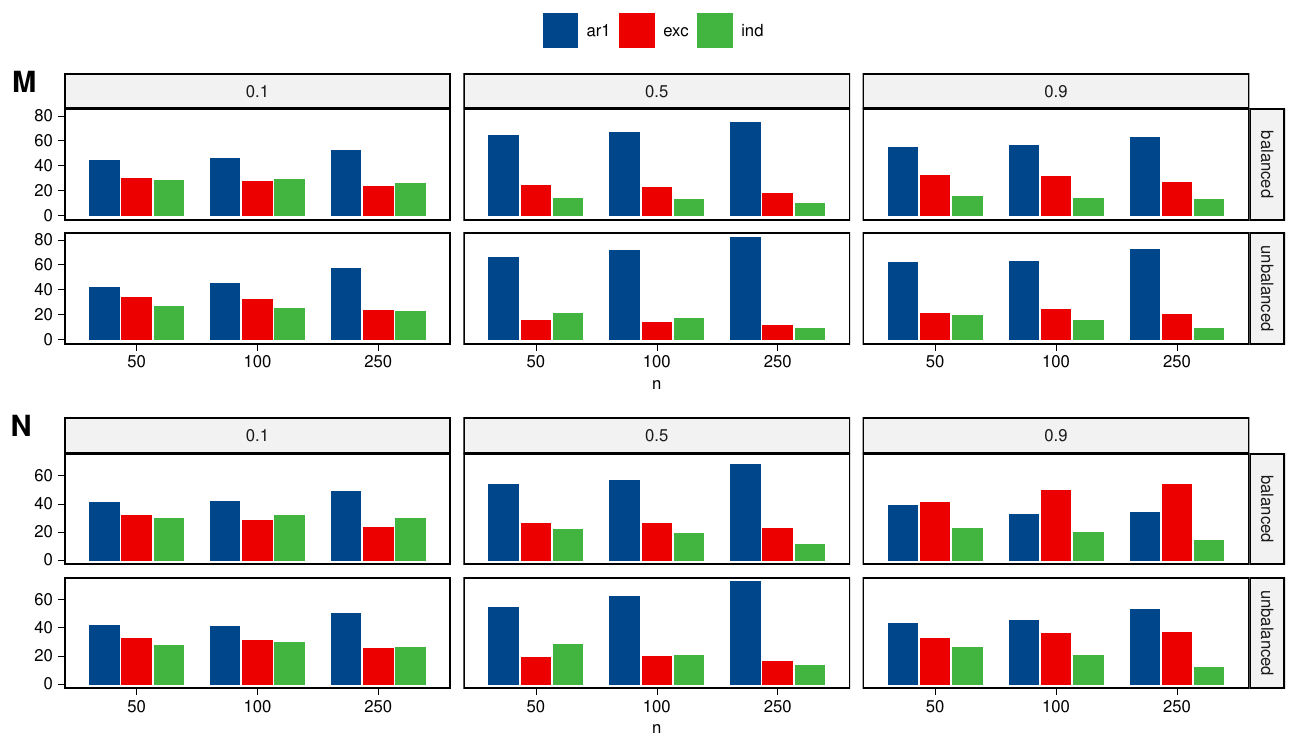}
  \caption{Percentage of working correlation structures selected by the $\boldsymbol{\QIC}$ criterion with respect to the sample
  size $n\in(50,  \ 100,  \ 250),$ the degree of correlation $\rho \in (0.1, \ 0.5, \ 0.9),$ the panel (balanced and unbalanced)
and the error term $\varepsilon\sim\mathcal{N}(0,1).$ Figures \textbf{M}-\textbf{N} represent the results for a location-shift and a location-scale-shift scenario, respectively.}\label{fig:QIC}
\end{figure}
\end{center}

\begin{center}
\begin{figure}[hbt!]
\includegraphics[width=1\linewidth]{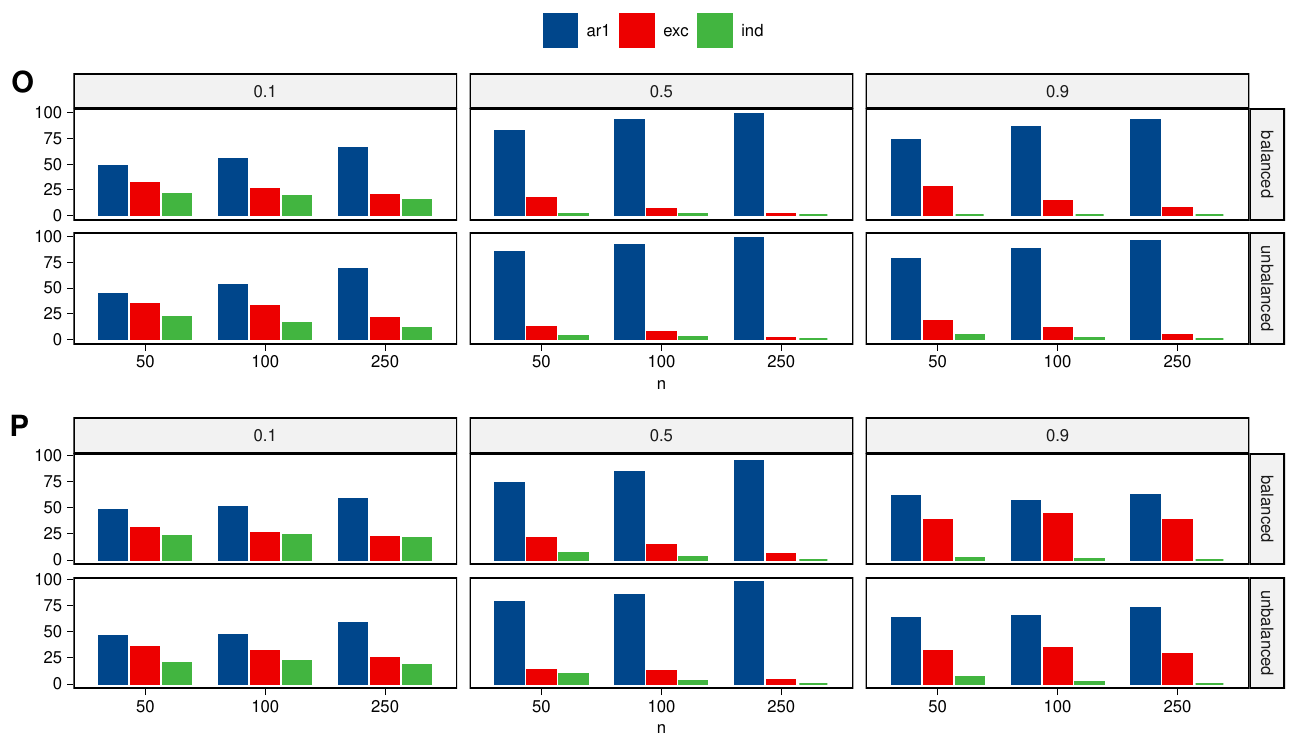}
  \caption{Percentage of working correlation structures selected by the $\boldsymbol{\CIC}$ criterion with respect to the sample
  size $n\in(50,  \ 100,  \ 250),$ the degree of correlation $\rho \in (0.1, \ 0.5, \ 0.9),$ the panel (balanced and unbalanced)
and the error term $\varepsilon\sim\mathcal{N}(0,1).$ Figures \textbf{O}-\textbf{P} represent the results for a location-shift and a location-scale-shift scenario, respectively.} \label{fig:CIC}
\end{figure}
\end{center}

\begin{center}
\begin{figure}[hbt!]
\includegraphics[width=1\linewidth]{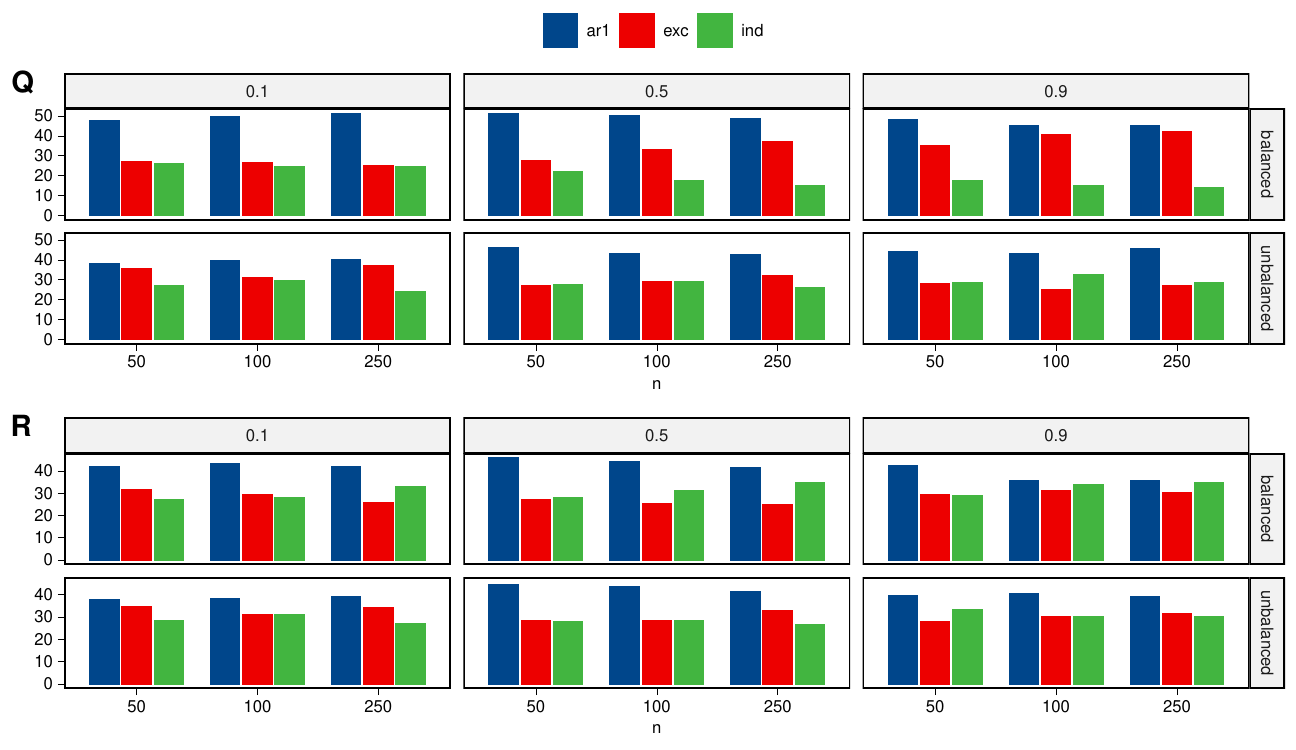}
  \caption{Percentage of working correlation structures selected by the asymmetric criterion $\boldsymbol{\asymQIC}$ with respect to the sample
  size $n\in(50,  \ 100,  \ 250),$ the degree of correlation $\rho \in (0.1, \ 0.5, \ 0.9),$ the panel (balanced and unbalanced)
and the error term $\varepsilon\sim\mathcal{N}(0,1).$ Figures \textbf{Q}-\textbf{R} represent the results for a location-shift and a location-scale-shift scenario, respectively.} \label{fig:asymQIC}
\end{figure}
\end{center}

\begin{center}
\begin{figure}[hbt!]
\includegraphics[width=1\linewidth]{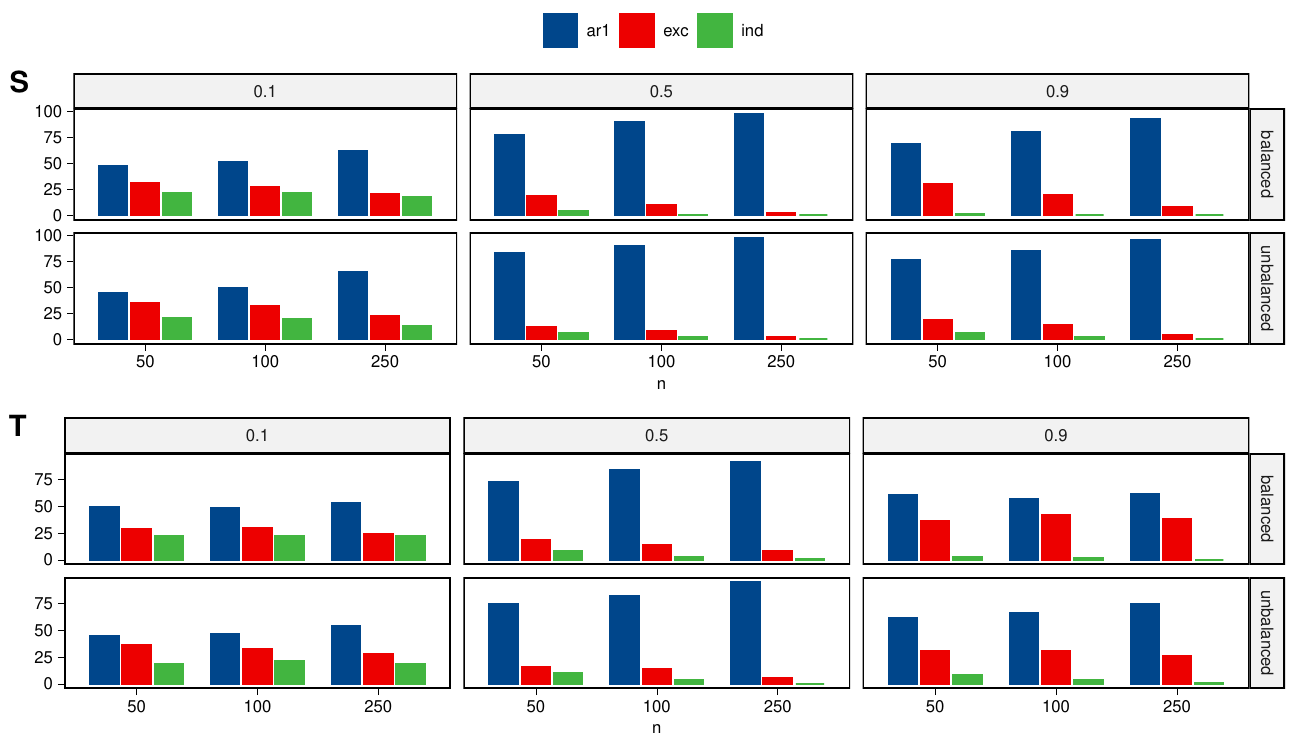}
  \caption{Percentage of working correlation structures selected by the asymmetric criterion $\boldsymbol{\asymCIC}$ with respect to the sample
  size $n\in(50,  \ 100,  \ 250),$ the degree of correlation $\rho \in (0.1, \ 0.5, \ 0.9),$ the panel (balanced and unbalanced)
and the error term $\varepsilon\sim\mathcal{N}(0,1).$ Figures \textbf{S}-\textbf{T} represent the results for a location-shift and a location-scale-shift scenario, respectively.} \label{fig:asymCIC}
\end{figure}
\end{center}

We assessed the goodness of fit of the GEEE model according to the selection criteria previously defined. We presented in \textbf{Figure \ref{fig:QIC}} and \textbf{Figure \ref{fig:CIC}} the percentage of selection of each working correlation structure according to the $\QIC$ and $\CIC$ measures, respectively. We also presented in \textbf{Figure \ref{fig:asymQIC}} and \textbf{Figure \ref{fig:asymCIC}} the percentage of selection according to the asymmetric measures, $\asymQIC$ and $\asymCIC,$ respectively. We observed that the $\QIC$ criterion selected the $\AR$ correlation structure as the true correlation structure most of the time. However, this observation was more pronounced with the $\CIC$ selection criterion. Indeed, the percentage of selection of the true correlation structure $(\AR)$ was greater than $50\%$ with the $\CIC$ measure, in all the scenarios, \textbf{Figure \ref{fig:asymCIC}}. 

The $\asymQIC$ and the $\asymCIC$ measures also selected in the majority of cases the true correlation structure $(\AR).$ The asymmetric selection criteria performed slightly better than the classical selection criteria, showing that, they could be a good alternative in the presence of heteroscedasticity in the data. Similar results were observed for the Student and  Chi-Squared distribution. These results can be found in the \textbf{supplementary material}.

Overall, the simulation results showed that our GEEE model yielded similar statistical properties as the classical GEE model. Indeed, the simulation results showed that our GEEE model was unbiased, consistent and highly efficient even with a misspecification of the true correlation structure \citep{LiangZeger1986}. Furthermore, our GEEE model naturally extended the various GEE working correlation structures, which offered flexibility in the specification of the random error correlation structure while accounting for the heteroscedasticity in the data. In the next section we highlighted the usefulness of our $\GEEE$ model in capturing the heteroscedasticity through the analysis of the labor pain dataset.

%% file: 5application.tex
\section{Application}\label{Application_geee}

In this section, we apply the $\GEEE$ model to estimate the reported labor pain score from the labor pain dataset \citep{davis1991}. The labor pain dataset is from a clinical trial that compares two treatments for maternal pain relief during labor. The dataset consists of 83 women in labor with 43 women randomly assigned to the treatment group and 40 women to the placebo group. The response variable (labor pain score) is a self-reported score measured every 30 min on a 100-mm line, where 0 means no pain and 100 means extreme pain. A nearly monotone pattern of missing data is found for the response variable, and the maximum number of measurements per woman is six. 

\begin{center}
\begin{figure}[hbt!]
\includegraphics[width=1\linewidth]{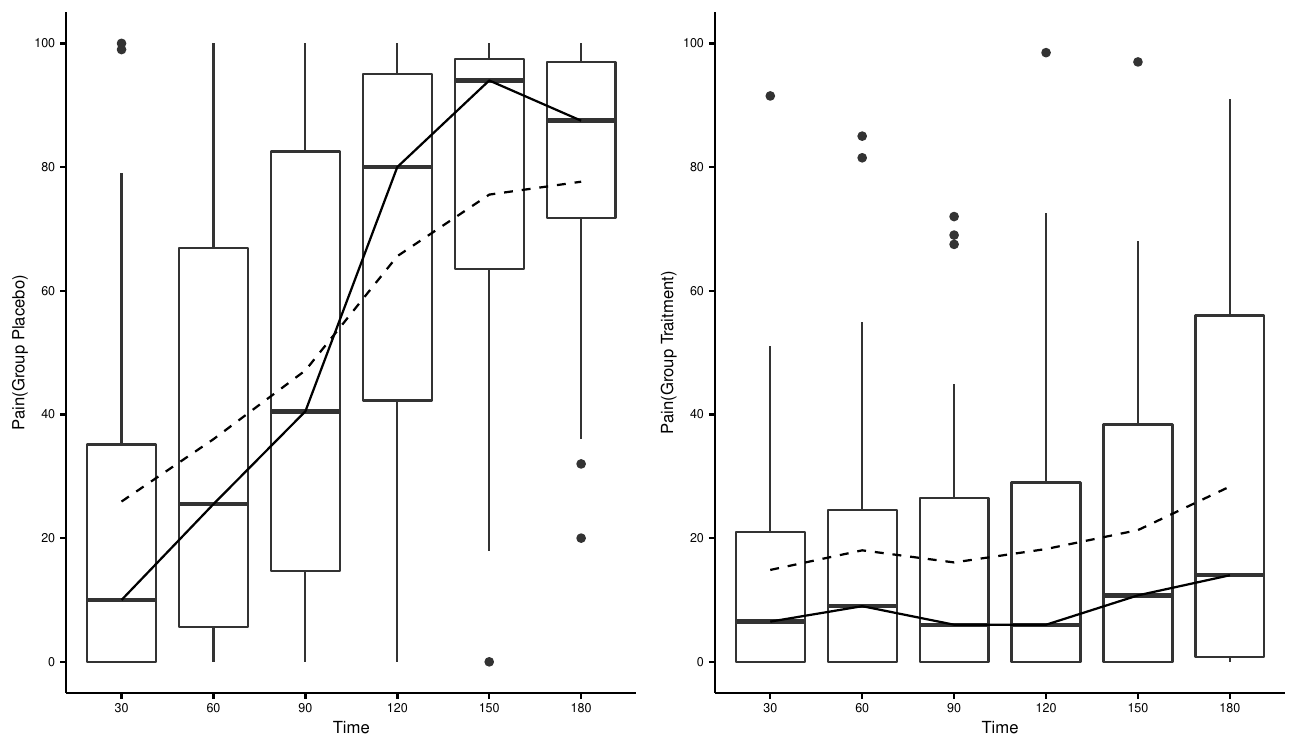}
  \caption{Boxplot of the labor pain score for the placebo and the pain medication groups. The
solid lines connect the medians and the dashed lines connect the means. }\label{fig:bplot}
\end{figure}
\end{center}

We plotted in \textbf{Figure \ref{fig:bplot}} the response distribution (boxplot) according to time and treatment. At first glance, the plots showed a time dependence of the labor pain distribution and a nonlinear trend of the mean as well as the median of the labor pain particularly in the placebo group.The shape of the boxplots and the varying position of the mean/median suggested that the magnitude and the sign of the skewness of the labor pain distribution changed over time for the placebo group. We observed that the distance between the second quartile and the median is greater than the distance of the median from the first quartile during the first 90 minutes before becoming smaller in the second half of the time. This suggested that the skewness of the pain response was positive before turning negative after 90 minutes. While the skewness for the pain medication group was always positive.  

Furthermore, the dataset consists of repeated measurements pertaining to the same subject. In such an occasion, a longitudinal model is more convenient and the GEE model could be a good candidate. Indeed, in his paper, \citet{davis1991} applied the semiparametric GEE model to estimate, as he called it, the extremely non-normal labor pain response. Unfortunately, the GEE model, although it models correctly the correlation structure of the random error, is not equipped to properly model the skewness of the heavy-tailed labor pain distribution. In contrast, the $\GEEE$ model by estimating the conditional expectiles of the labor pain distribution is able to account for the skewness of the heavy-tailed labor pain distribution. Hence, we apply the $\GEEE$ model to study the medication effect on the expectiles of the self-reported pain score distribution:

\begin{equation*}
    \mu_{\tau}(\boldsymbol{y}| \boldsymbol{X}) = \boldsymbol{X}\boldsymbol{\beta}_{\tau},
\end{equation*}

where the parameter $\boldsymbol{\beta}_{\tau}$ depends on $\tau.$ Through the estimation of the conditional expectiles, the $\GEEE$ model allows a detailed and comprehensive study of the medication effects on the pain score distribution. Additionally, the $\GEEE$ model offers a variety of correlation structures to flexibly model the within-subject dependence. This represents a great advantage of our models with respect to the GEE model, particularly when the response distribution is skewed or heavy-tailed. Notice that the $\GEEE$ model makes no assumptions about the distribution of the residual error.

To account for the time dependence of the treatment effect, we included an interaction term in the linear model. Furthermore, as suggested by the plots in \textbf{Figure \ref{fig:bplot}}, we also tested a square and a cubic term for the time variable and their interactions with the treatment. We presented here the result of the linear model including a square term for the time variable and its interaction with the treatment. The corresponding $\GEEE$ model is specified as:

\begin{equation}\label{eq:labor_pain_square}
\mu_{\tau}(y_{it}|R_i, T_{it}) = \beta_{0\tau} + \beta_{1\tau} R_i + \beta_{2\tau} T_{it} + \beta_{3\tau} R_i T_{it}
+ \beta_{4\tau} T_{it}^2 + \beta_{5\tau} R_i T_{it}^2,
\end{equation}

where \(R_i\) is the treatment variable \(0\) for the placebo group and $1$ for the medication group. The variable \(T_{it}\) is the measurement time divided by \(30\) min. We estimated the $\GEEE$ model at 9 expectiles $(0.1,\ 0.2,\ 0.3,\ 0.4,\ 0.5,\ 0.6,\ 0.7,\ 0.8,\ 0.9)$ and tested three working correlation structures: $\Ind, \ \Exc, \ \AR.$ Although the focus is more on the right tail of the labor pain distribution (women that are more sensitive to pain than average), it is still useful to have a detailed results for a complete and comprehensive analysis.
We relied on the $\CIC$ and the $\asymCIC$ criterion to select among the competing correlation structures.

The different $\GEEE$ working correlation structures produced comparable estimates and the $\CIC$ and $\asymCIC$ measures led to the selection of the $\GEEE$ model with the $\AR$ working correlation structure. The selection of the $\AR$ correlation model was consistent with the structure of the correlated data. Indeed, the repeated data was uniformly spaced in time and the correlation of the reported pain was stronger for adjacent measurements than for distant ones. 

\begin{center}
\begin{figure}[hbt!]
\includegraphics[width=1\linewidth]{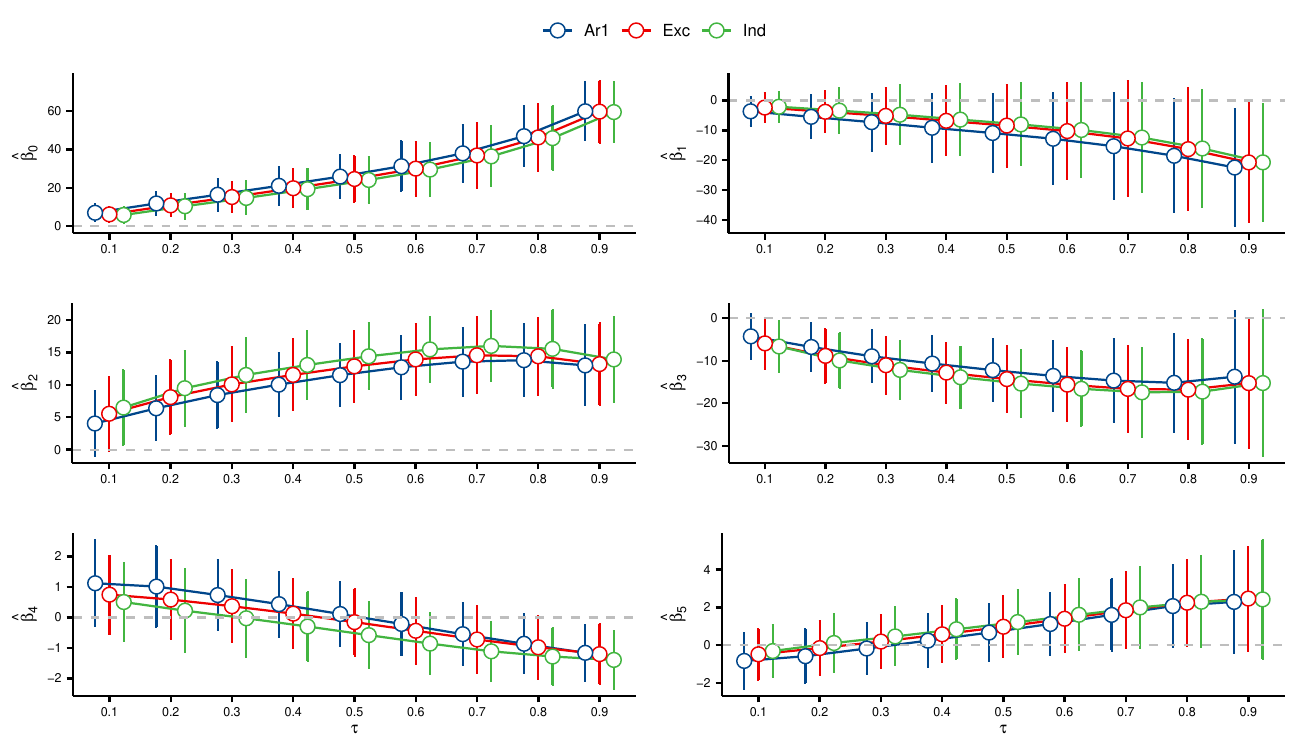}
  \caption{Parameter estimates and 95\% confidence interval of the $\GEEE$ independent, exchangeable and $\AR$ correlation models at
9 expectiles, $\tau\in(0.1,\ 0.2,\ 0.3,\ 0.4,\ 0.5,\ 0.6,\ 0.7,\ 0.8,\ 0.9).$ Note that the classical $\GEE$ corresponds to the $\GEEE$ model with $\tau=0.5.$  }\label{fig:est_geee}
\end{figure}
\end{center}

The result of the estimated parameters and their \(95\%\) confidence interval is presented in \textbf{Figure \ref{fig:est_geee}}. It is noticeable that the regressor effects vary from one expectile to another, and from one region to another of the response distribution. For example, the effect size of the estimated coefficient $\widehat{\beta}_4$ differs depending on whether one is in the center (null effect), on the left tail (positive effect), or on the right tail (negative effect) of the response distribution. This shows that the $\GEEE$ model, through the estimation of the conditional expectiles, offers a complete overview of the heterogeneous regressor effects on the response distribution, while correctly specifying the correlation structure of the random noise vector for the repeated measurements ($\AR$ working correlation structure). Whereas, the classic $\GEE$ model summarizes the complex relationship between the response and the regressors by estimating their effects on the mean of the response variable.

We observe that the effect size of the intercept estimate $(\widehat{\beta}_0)$ is significantly increasing across $\tau.$ Whereas, the effect size of the parameter estimate $\widehat{\beta}_1$ is rapidly decreasing and is not statistically significant at $(5\%)$ level, suggesting that the baseline pain does not globally differ between the placebo and the treatment group. 

The evolution of the difference between the placebo and the medication effects over time can be formalized as:

\begin{equation}\label{eq:labor_pain_square_geee}
    \mu_{\tau}(y_{it}|R_i=1)-\mu_{\tau}(y_{it}|R_i=0)= \beta_{1\tau} + \beta_{3\tau} T_{it} + \beta_{5\tau} T_{it}^2. 
\end{equation}

The effect size of the parameter estimate $\widehat{\beta}_3$ is negative and statistically significant at $5\%$ level. The effect size of the parameter estimate $\widehat{\beta}_5$ is subject to a sign shift and is significant at the extreme right tail of the distribution. The effect size of the parameter estimate $\widehat{\beta}_5$ has negative values and close to zero at the lower expectiles and at the expectiles close to the center $(\tau=0.5)$ and then its values turn positive at the higher expectile where it seems to be significant at the $5\%$ level. We observe that the effects of the three parameter estimates are simultaneously significant at the higher expectiles $(\tau = 0.8, \ 0.9),$ of the pain score distribution, \textbf{Figure \ref{fig:est_geee}}. In other words, the difference in pain between the two groups (placebo/treatment) is significant at the $5\%$ level for the fraction of women that is more sensitive to pain, which may suggest that the medication is effective. That said, these results should be interpreted with caution because the estimation of the tails of the distribution is more subject to sample size.

\begin{center}
\begin{figure}[hbt!]
\includegraphics[width=1\linewidth]{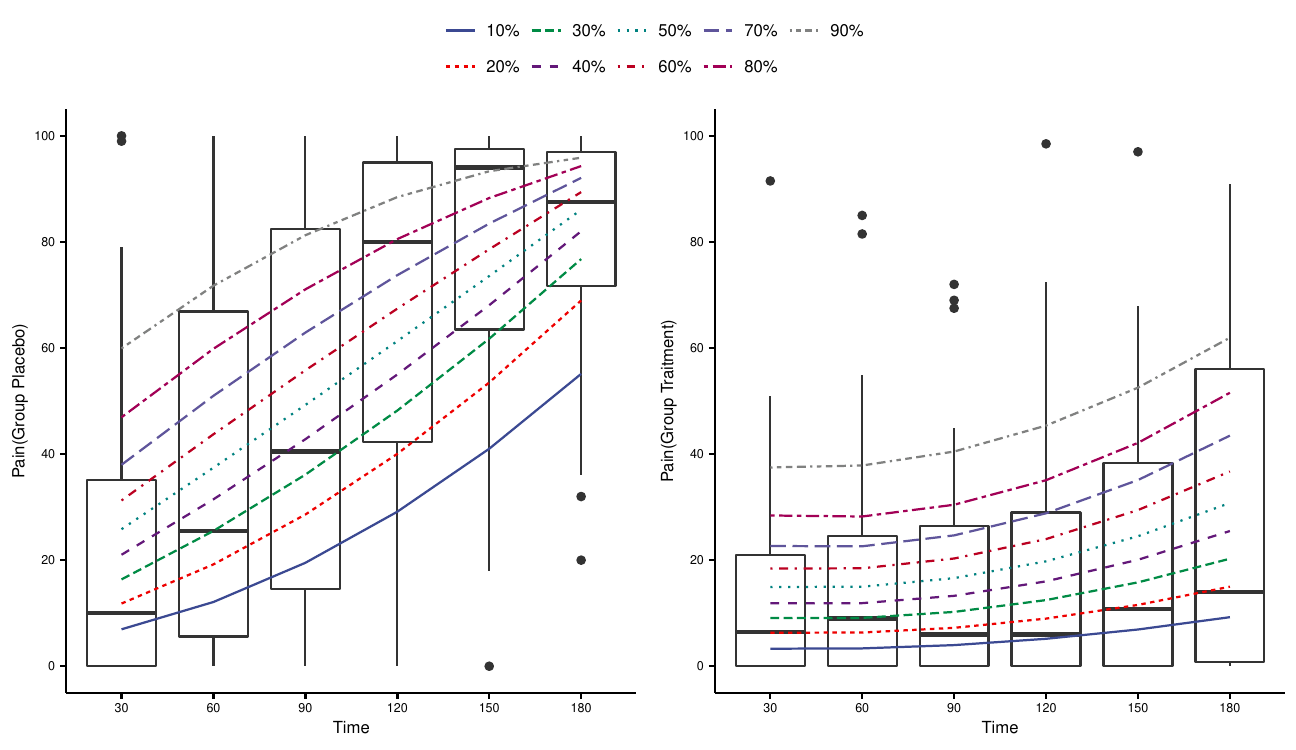}
  \caption{Boxplot of the labor pain score and the estimated curves of the expectile functions $(\tau\in(0.1,\ 0.2,\ 0.3,\ 0.4,\ 0.5,\ 0.6,\ 0.7,\ 0.8,\ 0.9))$ for the placebo group and the pain medication group using the $\GEEE$ and $\AR$ correlation model. Note that the classical $\GEE$ corresponds to the $\GEEE$ model with $\tau=0.5.$}\label{fig:pred_gee}
\end{figure}
\end{center}

The combined effect of the predictors in the placebo group and the treatment group is shown in \textbf{Figure \ref{fig:pred_gee}}. The results show that the curves of the conditional expectile functions capture well the skewness displayed by the boxplots of the labor pain variable. Whereas, the estimated curve of the $\GEE$ model alone does not account for these properties of the labor pain distribution. Indeed, as shown in \textbf{Figure \ref{fig:pred_gee}}, the estimated curve of the $\GEE$ model, $\GEEE(\tau=0.5),$ alone is ineffective to capture the heterogeneity displayed by the data.

Interestingly, we observe two trajectories in the placebo group: the women that are more sensitive to pain than average $(\tau=0.5)$ and the portion of women that is less sensitive to pain than average, \textbf{Figure \ref{fig:pred_gee}}. The women that are more sensitive to pain report an increasing pain and at a higher rate than average. While the women that are less sensitive to pain report an increasing pain, but at a lower rate. This difference may suggest the presence of a placebo effect at the lower tail of the response distribution. On the other hand, in the medication group,  we observe that the labor pain score starts at a low level and increases slowly compared to the placebo group, suggesting that the medication is effective, \textbf{Figure \ref{fig:pred_gee}}. Moreover, we observe that the curvature of the pain curves in the treatment group is similar to that of the curves of women who report less pain than average in the placebo group. Another result in favor of the efficiency of the medication treatment derived from the $\GEEE$ model. Note that these results should be interpreted with caution because not all parameter estimates at all levels of the conditional pain distribution are significant at the $5\%$ level. Additionally, a large sample size is necessary to study the tails of the response distribution and ensure the stability of the estimates. 

In summary, we applied our $\GEEE$ model and the classical $\GEE$ to study the effectiveness of the pain treatment for women in labor. The application of the $\GEE$ model alone did not allow us to conclude on the effectiveness of the labor pain medication. Whereas, the application of the $\GEEE$ model indicated that there is a pain difference between the proportions of women that is more sensitive to pain (the extreme right of the tail of the distribution) in the placebo group compared to the treatment group, suggesting that the medication was effective for this segment of the population. Despite the limitations of the sample (self measured pain, sample size, missing data), the data analysis highlighted the main advantage of the $\GEEE$ over the classical $\GEE,$ which is the description of the conditional response distribution through the estimation of its expectiles, while correctly specifying the correlation structure of the random noise vector for the repeated measurements.



%% file: 6conclusion.tex
\section{Conclusion}\label{Conclusion_geee}

We combined the weighted asymmetric least squares regression $(\ER)$ and the generalized estimating equations $(\GEE)$ to derive a new class of estimators, which we call the $\GEEE$ model. The $\GEEE$ estimates the conditional expectiles of the response distribution, which make it possible to capture regressor effects in the location, scale, and shape of the response distribution. This approach is useful in the presence of heteroscedasticity or when the response variable is asymmetric, skewed, or heavy-tailed. Additionally, the $\GEEE$ model offers a variety of correlation structures to flexibly model dependence among the observations, without making any assumption about the random error distribution.  

In this paper, we derived the asymptotic properties of the $\GEEE$ estimator and proposed a heterogeneous, consistent, and robust estimator of its variance-covariance matrix. The $\GEEE$ model is computationally efficient and easy to implement. See our GitHub for a free R package that simplifies the implementation (\url{github.com/AmBarry/expectgee}). We evaluated the finite sample properties of the $\GEEE$ model and the simulation results displayed its favorable qualities under various scenarios. We applied the $\GEEE$ model to the longitudinal clinical trial to evaluate the performance of two treatments (placebo/medication) with a skewed and heavy-tailed response variable. We showed that by estimating different expectiles of the response distribution, the $\GEEE$ model provides more information than the GEE model (about the relationship between the regressor and the response variables). 

Our $\GEEE$ model enhances the classical $\GEE.$ It allows the $\GEE$ model to account for the presence of heteroscedasticity in the data and capture regressor effects in the location, scale, and shape of the response distribution. Since $\GEE$ is a popular and useful statistical model for analyzing correlated data, we believe our $\GEEE$ model will be useful to researchers interested in analyzing longitudinal data. 

In terms of limitations: like the GEE model, the $\GEEE$ model is sensitive to outliers. We recommend using the numerous outlier detection tools available to reduce their influence prior to data analysis \citep{chatterjee_influential_1986}. In future work, it might be worthwhile to extend the $\GEEE$ model to account for missing data in longitudinal data analysis. Other complex longitudinal models that could benefit from the expectile regression model include linear mixed, logistic GEE, and count GEE.